\documentclass[a4paper,11pt]{article}
\pdfoutput=1
\usepackage{graphicx}
\usepackage{bm}
\usepackage{verbatim}
\usepackage{amsfonts}
\usepackage[latin1]{inputenc}
\usepackage{graphicx}
\usepackage{amssymb}
\usepackage{bm}
\usepackage{color}
\usepackage{ulem}
\usepackage{ulem}
\usepackage{float}
\usepackage{amsmath}
\usepackage{amsfonts}
\usepackage{dcolumn}
\usepackage{bm}

\def\be{\begin{equation}}
\def\ee{\end{equation}}
\def\ba{\begin{eqnarray}}
\def\ea{\end{eqnarray}}
\newcommand{\overbar}[1]{\mkern 1.5mu\overline{\mkern-1.5mu#1\mkern-1.5mu}\mkern 1.5mu}

\usepackage{jcappub} 

\title{Probing stochastic inter-galactic magnetic fields using blazar-induced gamma ray halo morphology}
\author{Francis Duplessis$^*$ and Tanmay Vachaspati$^{*\dag}$}
\affiliation{
	$^*$Physics Department, Arizona State University, Tempe, AZ 85287, USA. \\
	$^\dag$Maryland Center for Fundamental Physics, University of Maryland, 
	College Park, MD 20742, USA.}

\emailAdd{fdupless@asu.edu}
\emailAdd{tvachasp@asu.edu}

\abstract{
	Inter-galactic magnetic fields can imprint their structure on the morphology of blazar-induced gamma ray 
	halos. 
	We show that the halo morphology arises through the interplay of the source's jet and a two-dimensional 
	surface dictated by the magnetic field. 
	Through extensive numerical simulations, we generate mock halos created by stochastic magnetic fields 
	with and without helicity, and study the dependence of the halo features on the properties of the magnetic
	field.
	We propose a sharper version of the Q-statistics and demonstrate its sensitivity to the magnetic field
	strength, the coherence scale, and the handedness of the helicity. We also identify and explain a new 
	feature of the Q-statistics that can further enhance its power.}

\begin{document}

\maketitle
\flushbottom
\section{Introduction}

Multiple analyses of observed gamma rays~\cite{Neronov:1900zz,Ando:2010rb,Essey:2010nd,Tashiro:2013ita,
Chen:2014qva,Chen:2014rsa,Finke:2015ona} provide growing evidence for the existence
of inter-galactic magnetic fields (for reviews see \cite{Brandenburg:2004jv,Durrer:2013pga}). 
The existence of such magnetic fields poses new questions for cosmology and probably also for particle 
physics~\cite{Wagstaff:2014fla,Vachaspati:2016xji}. In addition, a primordial magnetic field can play
an important role during structure formation in the universe and could
help us understand the ubiquity of magnetic fields in astrophysical bodies.

A critical challenge at this stage is to sharpen observational techniques so that we can better observe and 
measure inter-galactic magnetic fields. Of the various probes of inter-galactic magnetic fields, blazar-induced 
gamma ray cascades hold certain key advantages. The gamma ray cascades originate in the voids
in the large-scale structure and are mostly immune to complications of a noisy environment. The cascade develops
in a relatively small spatial volume and hence is a local probe of the magnetic field in the voids. This
is distinct from other methods, such as the Faraday rotation of the cosmic microwave background
polarization, that probe an integrated measure of the magnetic field. Gamma ray cascades are also highly
sensitive probes and can trace very weak cosmological magnetic fields.

In this paper we study the effect of stochastic inter-galactic magnetic fields on blazar induced gamma 
ray halos and some results overlap with those of Refs.~\cite{Elyiv:2009bx,Long:2015bda,AlvesBatista:2016urk,
Broderick:2016akd,Fitoussi:2017ble}.
The cascade process is complicated and all analyses use some simplifying assumptions. 
For example, the analysis in Ref.~\cite{Long:2015bda} only considered
non-stochastic magnetic field configurations. Other simplification schemes, such as the ``large
spherical observer'' method employed in Ref.~\cite{AlvesBatista:2016urk}, transport arrival 
directions of gamma rays for distant observers to a single Earth-bound observer. This
technique is certainly useful to study spectral properties of the cascade, but there is a
danger that it loses or shuffles the spatial information of gamma ray arrival directions that 
is crucial for morphological studies. Our focus is on the effect of stochastic magnetic
fields that are statistically isotropic and with or without helicity. So we carefully analyse
the spatial information of the gamma rays that is useful for deducing properties of the
magnetic field but, for the present, we only include an approximate description of the
cascade development.

An important helpful concept that we develop in this paper is that of the ``PP surface''
(see Sec.~\ref{sec:morph}). This spatial surface holds the key to halo morphology and many 
of the features that we see in our simulations can be understood in terms of the shape 
of the PP surface and its intersection with the blazar jet.

We have applied a refined version of the Q-statistics first proposed in Ref.~\cite{Tashiro:2013bxa} 
to study the morphology of halos. Our results show that this statistic can successfully extract
the helicity of the magnetic field. Our simulations also reveal that the plot of $Q(R)$, where
$R$ is a variable that will be explained below, has an additional bump. We are able to show
that this bump is a genuine feature of the Q-statistic and explain it in terms of properties of the 
PP surface in Sec.~\ref{sec:bump}. Thus this extra feature of $Q(R)$ may become an
observational tool in future.

We give some background information in Sec.~\ref{background}, discuss our simulation
techniques in Sec.~\ref{sec:morph}, discuss features of the halo in Sec.~\ref{halofeatures},
introduce the Q-statistic in Sec.~\ref{Qanalytical} and apply it to stochastic
fields in Sec.~\ref{Qstochastic}. As mentioned above, we discuss the bump feature in
$Q(R)$ in Sec.~\ref{sec:bump}. 
We summarize our conclusions in Sec.~\ref{conclusions}. Our stochastic magnetic field
generation scheme is described in Appendix~\ref{Bgeneration}.

\section{Blazar Halos from an Intergalactic Magnetic Field}
\label{background}

TeV photons from blazars induce electromagnetic cascades through pair production with the extragalactic background light,
$
\gamma_{_\text{EBL}}\gamma_{_\text{TeV}}\rightarrow e^+ e^-.
$
In the presence of a magnetic field, the charged leptons follow spiral paths as they propagate and lose energy
due to inverse Compton scattering with the cosmic microwave background (CMB) photons.
The up-scattered CMB photons have gamma ray energies and produce extended halos around the direction
of the blazar. In this section we briefly discuss the formation of the halo under simplifying assumptions.

Consider a blazar located at the origin of our coordinate system described by the unit basis vectors 
$\hat{\mathbf{x}},\hat{\mathbf{y}},\hat{\mathbf{z}}$.
We choose $\hat{\mathbf{z}}$ so that Earth is located at $\mathbf{r}_{_\text{E}}=-d_s \hat{\mathbf{z}} $ 
where $d_s$ is the comoving distance to the source,
\be
d_s=\frac{1}{a_0H_0}\int_0^{z_s}\frac{1}{\sqrt{\Omega_m(1+z)^3+\Omega_\Lambda}}dz\simeq \frac{z_s}{0.22}\text{Gpc}.
\ee
To perform the integral, we have used $\Omega_\Lambda\approx 0.69$, $\Omega_m\approx 0.31$, $H_0\approx 0.67h$ 
as found in Ref.~\cite{Ade:2015xua} and we have also assumed that $z_s \ll 1 $ and used natural units so that $c=1$. 
For all the simulations in this paper we will choose $d_s=1~{\rm Gpc}$.

The blazar will typically emit photons in a collimated jet which we approximate to be a conical region with half-opening 
angle $\theta_{\text{jet}} \approx 5^\circ$.
The energy $E_{\gamma 0}$ of these photons must lie above some threshold of about a TeV if they are to produce 
an electron-positron pair from interaction with the Extragalactic Background Light (EBL).
Due to the opacity of the EBL, the TeV photons will travel a mean free path (MFP) determined by the pair production 
cross section $\sigma_{\gamma\gamma}$ and the number density of the EBL photons $n_{_\text{EBL}}$,
\be
D_{\gamma 0}=\langle \sigma_{\gamma\gamma} n_{_\text{EBL}} \rangle^{-1}\simeq (80 \text{Mpc}) \frac{\kappa}{(1+z_{\gamma\gamma})^2}\Big(\frac{10\text{TeV}}{E_{\gamma 0}}\Big),
\ee
We have assumed that $n_{_\text{EBL}}\propto (1+z_{\gamma\gamma})^{-2}$ to approximate the MFP in the final 
equality~\cite{Neronov:2009gh}. Following \cite{Long:2015bda} we will set
$\kappa=1 $ as this dimensionless constant is estimated to lie in the range of $0.3< \kappa < 3$. 
The comoving distance from the source to the pair production event is given by
$D^c_{\gamma 0}=(1+z_{\gamma\gamma})D_{\gamma 0}$.

The redshift of the produced lepton pairs will depend on the relative position of the leptons to the source. 
Since $D^c_{\gamma 0}\ll d_s$, 
we make the approximation $z_{\gamma\gamma}\approx z_s$ and we can write,
\be\label{TeVcoMFP}
D^c_{\gamma 0}\simeq (80 \text{Mpc}) \frac{\kappa}{(1+z_{s})}\Big(\frac{10\text{TeV}}{E_{\gamma 0}}\Big).
\ee

The energy of each of the produced leptons will be $E_e\approx E_{\gamma 0}/2$. 
These leptons are expected to travel a distance $D_e$ before losing most of their energy through inverse 
Compton (IC) cooling which occurs
by upscattering CMB photons. The cooling distance is
\be
D_e=\frac{3m_e^2}{4\sigma_T U_{CMB}E_e}
\simeq (31 \text{ kpc})\Big(\frac{5\text{ TeV}}{E_e}\Big)\Big(\frac{1.22}{1+z_{\gamma\gamma}}\Big)^4
\label{eq:De}
\ee
where $\sigma_T=6.65\times 10^{-25} \text{cm}^2$ is the Thomson scattering cross section and 
$U_{CMB}(z_{\gamma\gamma})\simeq (0.26 ~\text{eV}/\text{cm}^3)(1+z_{\gamma\gamma})^4$
is the CMB energy density. Note that we can assume the whole cascade development
happens around redshift $z_{\gamma\gamma}$ as $D_e\ll D_{\gamma 0}$.
At that redshift, the average energy of a CMB photon is 
\be
E_{CMB} \simeq  ( 6\times 10^{-4}\text{ eV})(1+z_{\gamma\gamma}),
\ee
which implies that, from energy conservation, the upscattered photons will have energy 
\be\label{igamogamenergy}
E_\gamma=\frac{4}{3}E_{CMB}\frac{E_e^2}{m_e^2}\simeq (77 \text{ GeV})\Big(\frac{E_{\gamma 0}}{10~\text{TeV}}\Big)^2.
\ee
As the lepton propagates, it upscatters $\approx (10 {\rm TeV})/(10 {\rm GeV}) \sim 10^3$ photons, and produces a
gamma ray cascade in the 1-100~GeV range if the initial gamma ray had an energy of a few TeV. 
Clearly not every photon upscattered by the leptons will reach Earth.
Those that do must come from a set of events that satisfy a set of three constraints given in 
Ref.~\cite{Long:2015bda} that we now describe. 

After pair production, the lepton's initial velocity will be almost parallel to the momentum of the parent photon
with a negligible deviation of order the inverse Lorentz boost factor $m_e/E_e\sim 10^{-6}$. Their subsequent 
trajectory will be determined by the magnetic field $\mathbf{B(x)}$ through the Lorentz force. If the magnetic
field is incoherent on length scales smaller than the cooling distance $D_e \sim 30~{\rm kpc}$, the lepton trajectories 
will be diffusive and this situation is much harder to analyze. So we focus on magnetic fields that are coherent on
scales that are much larger than $D_e$. Then the lepton trajectories are bent in an effectively constant magnetic
field and follow a helical trajectory with gyroradius
\be
R_L=R_{L0} |\mathbf{v}_\perp|,~~~\text{with } R_{L0}=\frac{E_e}{e|\mathbf{B}|},
\ee
which depends on the lepton's perpendicular velocity to $\mathbf{B}$, 
{\it i.e.} $\mathbf{v}_\perp=\mathbf{v}-(\mathbf{v}\cdot\mathbf{\hat{B}}) \mathbf{\hat{B}}$.

The quantity $2\pi R_{L0}$ is useful as it denotes the distance the lepton must travel in order to perform a full revolution.
The value of $R_{L0}$ is a function of the redshift as it depends on $|\mathbf{B}|$. 
For magnetic fields frozen in the plasma,
the field strength redshifts as $|\mathbf{B}|= B_0 (1+z_{\gamma\gamma})^2\approx B_0 (1+z_{s})^2$ where
$B_0$ is the magnetic field magnitude today. With
\be
R_{L0} \simeq 3.5\text{ Mpc}~
\Big(\frac{E_e}{5\text{ TeV}}\Big)\Big(\frac{B_0}{10^{-15} G}\Big)^{-1}\Big(\frac{1+z_s}{1.22}\Big)^{-2},
\ee
we can evaluate the ratio 
\be
\frac{D_e}{2\pi R_{L0}}\simeq 0.0106 ~
\Big(\frac{E_\gamma}{10\text{ GeV}}\Big)^{-1}\Big(\frac{B_0}{10^{-15} G}\Big)\Big(\frac{1+z_s}{1.22}\Big)^{-2}
\label{bending}
\ee
which determines the angular deflection of the leptons.

\begin{figure}
	\centering
\includegraphics[width=0.8\textwidth]{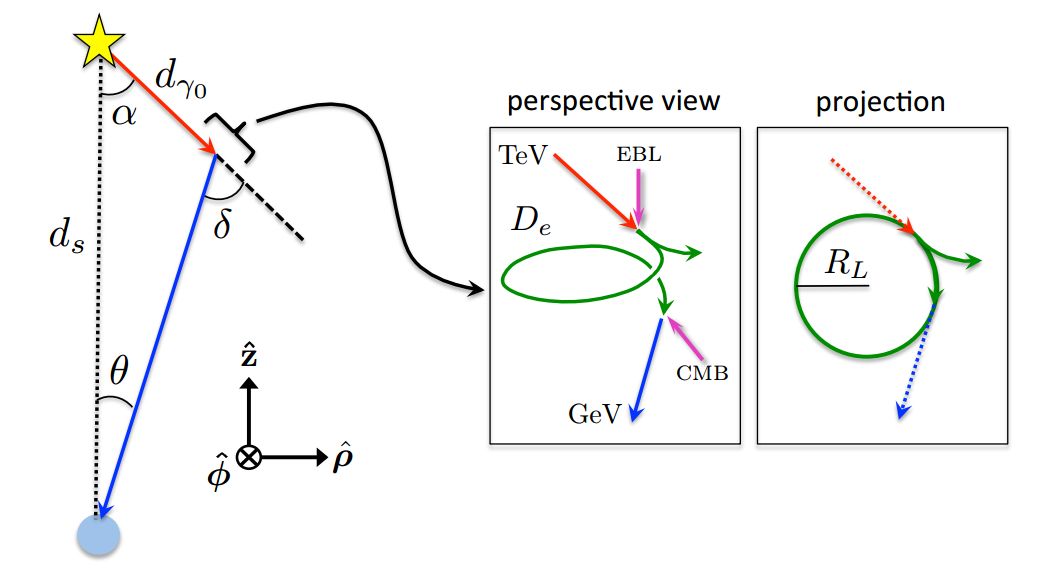}
\caption{A TeV photon emitted from a blazar travels a comoving distance of $d_{\gamma 0}$ before scattering 
off an EBL photon and pair producing leptons. The lepton trajectories are bent due to the local magnetic field over 
a very short distance compared to the distance to the source, $d_s$, and are shown in the insets. 
(The insets show huge bending whereas we have only considered magnetic fields that give small bending.)
Inverse-Compton scattering of a lepton and CMB photons results in a cascade
of GeV energy gamma rays arriving at Earth from the direction of the pair production.
[Sketch taken from \cite{Long:2015bda}.]}
\label{halocreation}
\end{figure}

As depicted in Fig.~\ref{halocreation}, let us
introduce the following angles: $\theta$ is the arrival angle of the GeV photon with respect to the source location, 
$\delta$ represents the angle between the upscattered photon and the TeV photon, $\alpha$ is the angle subtended 
by the TeV photon's momentum, $\mathbf{r}_E$ the vector from the source to Earth, and finally $\phi$ is the 
azimuthal angle in which the whole (planar) scattering process take place. We also introduce the polar vectors 
$\bm{\hat{\rho}}$ and $\bm{\hat{\phi}}$,
in the $\mathbf{\hat{x}},\mathbf{\hat{y}}$ plane. It is important to emphasize that the whole process occurs in a plane
to a good approximation because $D_e \ll D^c_{\gamma 0}, d_s$ and the length of the lepton trajectory can be ignored. 
Then there are only 3 points that are relevant (the source, the pair production point, and the observer) and they
always lie in a plane.

Applying the sine formula to the triangle in Fig.~\ref{halocreation} we get our first constraint

\be\label{const1}
d_s \text{sin}(\theta)=d_{\gamma 0} \text{sin}(\delta).
\ee
where $d_{\gamma 0}$ is the distance traveled by the TeV photon and is a random variable
drawn from a distribution that depends on the MFP $D^c_{\gamma 0}$ in Eq.~(\ref{TeVcoMFP}).
This is discussed in detail in Sec.~\ref{sec:morph}.

The bending angle $\delta$ is related to the distance traveled by the lepton through the local magnetic field which 
we write as $\mathbf{B}=B \mathbf{\hat{n}}_{||}$. We also decompose the lepton's initial velocity at time $t_i=0$,
$\mathbf{v}(t_i=0)=v_{||}\mathbf{\hat{n}}_{||}+v_\perp\mathbf{\hat{n}}_{\perp}$, where 
$\mathbf{\hat{n}}_{\perp}\cdot \mathbf{\hat{n}}_{||} = 0$. At some later time $t$ the velocity is
\be\label{leptonvel}
\mathbf{v}(t)=v_{||}\mathbf{\hat{n}}_{||}+v_\perp\text{cos}(\omega t)\mathbf{\hat{n}}_{\perp}\pm v_\perp\text{sin}(\omega t)(\mathbf{\hat{n}}_{\perp}\times \mathbf{\hat{n}}_{||}),
\ee
Here we introduced the angular frequency of the orbital motion $\omega=v_\perp/R_L=1/R_{L0}$ 
and the $+ ~ (-)$ sign refers to the positron (electron) trajectory. A CMB photon upscattered at time $t_{IC}$ will 
be directed along the lepton's trajectory and so the deflection angle of Figure \ref{halocreation} can be expressed as 
$\text{cos}(\delta)=\mathbf{\hat{v}}(0)\cdot\mathbf{\hat{v}}(t_{IC})$. 
Using Eq. (\ref{leptonvel}) we can derive the second constraint,
\be\label{const2}
1-\text{cos}(\delta)=\Big(1-(\mathbf{\hat{v}}(0)\cdot \mathbf{\hat{B}})^2\Big)\Big(1-\text{cos}(t_{IC}/R_{L0})\Big).
\ee
The time of inverse Compton scattering $t_{IC}$ is a stochastic variable. Given its value and
the magnetic field direction, the constraints determine the bending angle, $\delta$.

A single propagating lepton will be able to upscatter CMB photons towards Earth only at certain times
when the lepton's momentum is directed towards Earth. Photons upscattered at other times will not reach 
Earth and we can safely ignore them. The number of photons upscattered by a lepton is very large ($\sim 10^3$),
with mean deviation angles between the photons $\sim 10^{-3}\times 0.01$ (see Eq.~(\ref{bending})). This angle 
is large enough that we only expect $\sim 1$ of the cascade photons from any lepton to reach Earth. This allows 
us to adopt the strategy that we first select a value of $t_{IC}$ from an exponential probability
distribution as described in Sec.~\ref{sec:morph} and then solve the constraint equations to find all
TeV gamma rays from the blazar that upscatter CMB photons that reach Earth. For different values of
$t_{IC}$, different TeV gamma rays from the blazar will lead to observed photons.
In this way, we will be able to track the photons that arrive on Earth and not waste computational effort
on those that go elsewhere.

The third and final constraint is that the cascade gamma ray lies in the plane specified by $\mathbf{\hat{\phi}}$.
This requires that the Lorentz force in the azimuthal $\mathbf{\hat{\phi}}$ direction vanishes between the time of 
pair production and IC scattering. Namely the $\phi$ component of the impulse must vanish,
\be
J_\phi=\bm{\hat{\phi}}\cdot \mathbf{J}=
\bm{\hat{\phi}}\cdot\Big(\pm e\int_0^{t_{IC}} dt ~\mathbf{v}(t)\times \mathbf{B} \Big)=0 .
\ee
The impulse can be simplified by pulling out the assumed constant magnetic field of the integral and defining
\be
\mathbf{v}_{\text{avg}}=\frac{1}{t_{IC}}\int_0^{t_{IC}} dt ~\mathbf{v}(t).
\ee
The geometrical setup of Fig.~\ref{halocreation} forces $\mathbf{v}_{\text{avg}}$ to bisect the angle 
$\delta$ and therefore its unit vector can be written as
\be
\mathbf{\hat{v}}_{\text{avg}}=\text{sin}\big(\delta/2-\theta\big)\hat{\bm{\rho}} -\text{cos}\big(\delta/2-\theta\big)\mathbf{\hat{z}}.
\ee
Decomposing $\mathbf{B}$ as
\be
\mathbf{B}=b_\rho \hat{\bm{\rho}}+b_\phi \hat{\bm{\phi}}+b_z \mathbf{\hat{z}}
\ee
allows us to write
\be\label{const3}
\bm{\hat{\phi}}\cdot \mathbf{\hat{v}}_{\text{avg}}\times \hat{\mathbf{B}}=-b_\rho\text{cos}\big(\delta/2-\theta\big)-b_z\text{sin}\big(\delta/2-\theta\big)=0.
\ee

To summarize this section, Eqs.~(\ref{const1}), (\ref{const2}) and (\ref{const3}) are the constraints that need to 
be satisfied by the variables $(\theta,\delta,\phi)$ given 
a magnetic field realization and initial velocity of the TeV gamma ray (both of which depend on $(\theta,\delta,\phi)$), 
the source-observer distance ($d_s$), the distance to pair production ($d_{\gamma 0}$), and the photon 
upscattering time ($t_{IC}$).

\section{Halo Simulations}
\label{sec:morph}

For events that satisfy the constraints in Eqs.~(\ref{const1}), (\ref{const2}) and (\ref{const3}), an observer on Earth
will receive flux at a polar angle of $\theta$ from the line of sight (LoS) to the blazar and at an azimuthal angle $\phi$. 
Solving these constraints requires the use of numerical methods when considering general 
${\bf B}(\bf x)$ and when including the stochasticity in the propagating distances (PDs) of the initial gamma 
ray and pair produced leptons.

Therefore to simulate one observed photon, we supply the distance $d_{\gamma 0}$ traversed by 
some TeV gamma ray of energy $E_{\gamma 0}$ emitted from the source
before it pair produces leptons, one of which in turn travels a distance $c t_{IC}$ before emitting a photon of energy $E_\gamma$. Once these 4 values $d_{\gamma 0}$, $E_{\gamma 0}$, $c t_{IC}$ and $E_\gamma$,
are set and an ambient magnetic field is given, one can numerically solve the constraint equations
for $\theta,~\delta$ and $\phi$.  The process is repeated until $N$ (which we chose to be 1000 or 5000 per simulation) observed photons are simulated. This will create the halo that one would observe if the source was emitting isotropically.  For a source with a specific jet orientation we only retain the events whose initial TeV photons lie within the jet. These small number of events give us the observed halo that will be shown in our plots.

\begin{figure}[tbhp]
	\centering
	\includegraphics[width=0.48\textwidth]{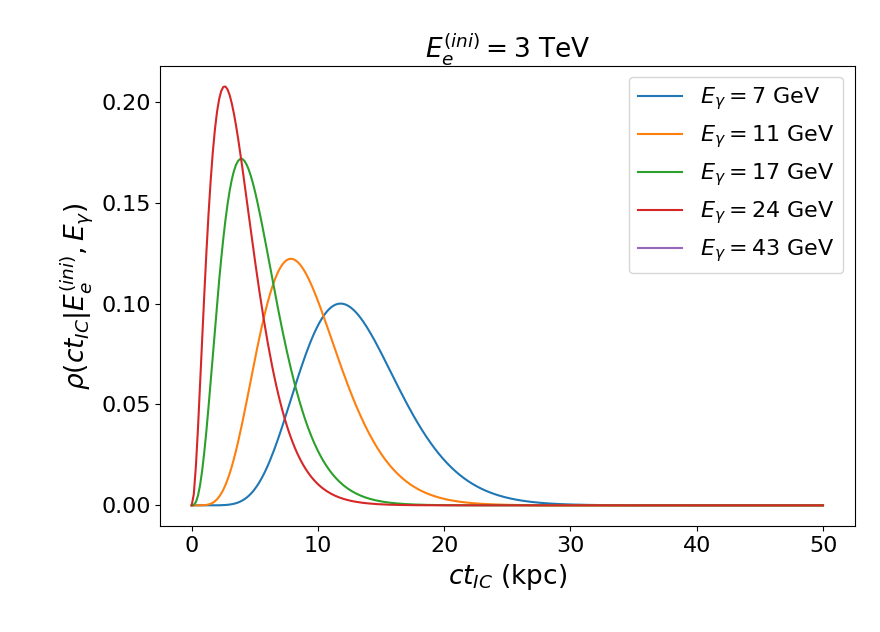}
	\includegraphics[width=0.48\textwidth]{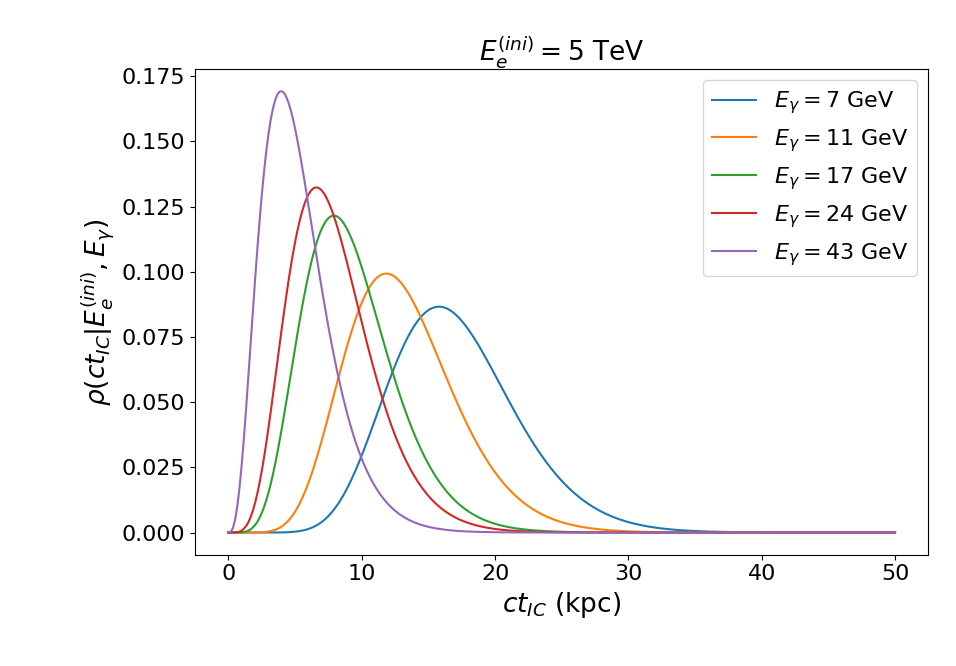}
	\caption{Examples of the probability distribution for the distance traveled by a lepton with initial energy 
	$E_e=E_{\gamma_0}/2$ before it upscatters a CMB photon to $E_{\gamma}$. These distributions are for 
	leptons evolving in the CMB light at a redshift of $z\approx 0.24$,
	corresponding to a source located at $d_s\approx 1$~Gpc from Earth. Note that the final distribution for 
	$E_\gamma=43$~GeV does not appear on the left as the lepton does not possess enough energy to 
	upscatter the CMB photons to these energies.}
	\label{distex}
\end{figure}

Let us go through the details regarding the generation of $d_{\gamma 0}$, $E_{\gamma 0}$, $c t_{IC}$ and $E_\gamma$.
We must supply some energy distribution for gamma rays emitted by blazars; for this  
we assume a power law spectrum \cite{Ackermann:FS2,TheFermi-LAT:2015ykq} and follow 
Ref.~\cite{Broderick:2016akd} by choosing a spectral index of $\Gamma\simeq 2.5$ that is characteristic 
of the TeV sources. This yields a spectrum given by, 
\be
  \frac{dN_{\gamma 0}}{dE_{\gamma 0}}\sim \Big(\frac{E_{\gamma 0}}{\text{TeV}}\Big)^{-2.5}.
\ee 
We shall also impose a $10$~TeV cutoff on the emitted photon energy.
The distance $d_{\gamma 0}$ traversed by a TeV gamma ray before turning into a pair of leptons is drawn from the exponential distribution,
\be
P[d_{\gamma 0}] = \frac{1}{D^c_{\gamma 0}(E_{\gamma 0})} e^{-d_{\gamma 0}/D^c_{\gamma 0}(E_{\gamma 0})} .
\label{xdist}
\ee
The resulting leptons will have energies $E_e$ given by Eq.~(\ref{igamogamenergy}) and they will upscatter numerous 
CMB photons along their trajectories. The mean free path between each scattering is 
given by $l_{\text{MFP}}=(n_{CMB}\sigma_T )^{-1}$, where $n_{CMB}$ is the number density of CMB photons.
The lepton loses energy with each scattering and subsequent scatterings lead to lower energy cascade gamma rays.
Hence we run Monte Carlo simulations to determine the distributions 
$P(c t_{IC},E_{\gamma}|E_e^{(\text{ini})})$,
giving us the probability that a lepton with initial energy $E_e^{(\text{ini})}$ upscatters a CMB photon to energy 
$E_\gamma$ after traveling a distance $d_e=c t_{IC}$. Examples of these distributions are shown in Fig.~\ref{distex}.
Note that $d_e$ will generally be much smaller than the cooling distance $D_e$.  Only events that lead to observed photons of energy between $E_{\rm min}=5~{\rm GeV}$ and $E_{\rm max}=50~{\rm GeV}$ will be retained as these are in energies of observational interest for the statistical analysis done in Sec.~\ref{Qanalytical}.

We will solve the constraint equations in a variety of magnetic field backgrounds, starting with
simple analytic configurations for illustration purposes, and then move on to the more realistic
case of stochastic, isotropic magnetic fields. Our procedure to generate stochastic, isotropic magnetic 
fields is described in Appendix~\ref{Bgeneration}.

As a warm up, and to compare our method with the results of Ref.~\cite{Long:2015bda},
we consider a source that radiates TeV photons isotropically in two different magnetic field
backgrounds. The first background,
\be
\label{constantB}
{\bf B} =B_0\Big(\text{cos}(\beta)\hat{\bf y}-\text{sin}(\beta) \hat{\bf z}\Big),
\ee
with $\beta=\pi/4$, is a uniform magnetic field pointing at an angle $\pi /4$ from the line of sight.
The second background is a maximally helical field
\be
\label{helB}
{\bf B} =B_0\Big(\text{sin}(2\pi z/\lambda) \hat{\bf x}+\text{cos}(2\pi z/\lambda)\hat{\bf y}\Big).
\ee
Here $\lambda$ is the coherence length of
the helical field.
We will take $d_s=1 ~{\rm Gpc}$, $B_0=10^{-14}~{\rm G}$ and $\lambda=500\text{ Mpc}$ as the prototypical 
values and eventually vary them one at a time to see their effect on the halo morphology.

Next we solve the constraint equations and determine the arrival directions 
$\theta$, $\phi$ for several different energies $E_\gamma$, for an isotropically emitting source.
The points located further away from the source direction 
usually corresponds to lower energy photons. This is expected since leptons that travel long distances 
(and hence allow for a large bending angle) will have already lost a lot of energy and upscatter less 
energetic photons. This behavior can be seen from the distribution shown in Fig~\ref{distex}.

\begin{figure}[tbhp]
  	\centering
  	\includegraphics[width=0.45\textwidth]{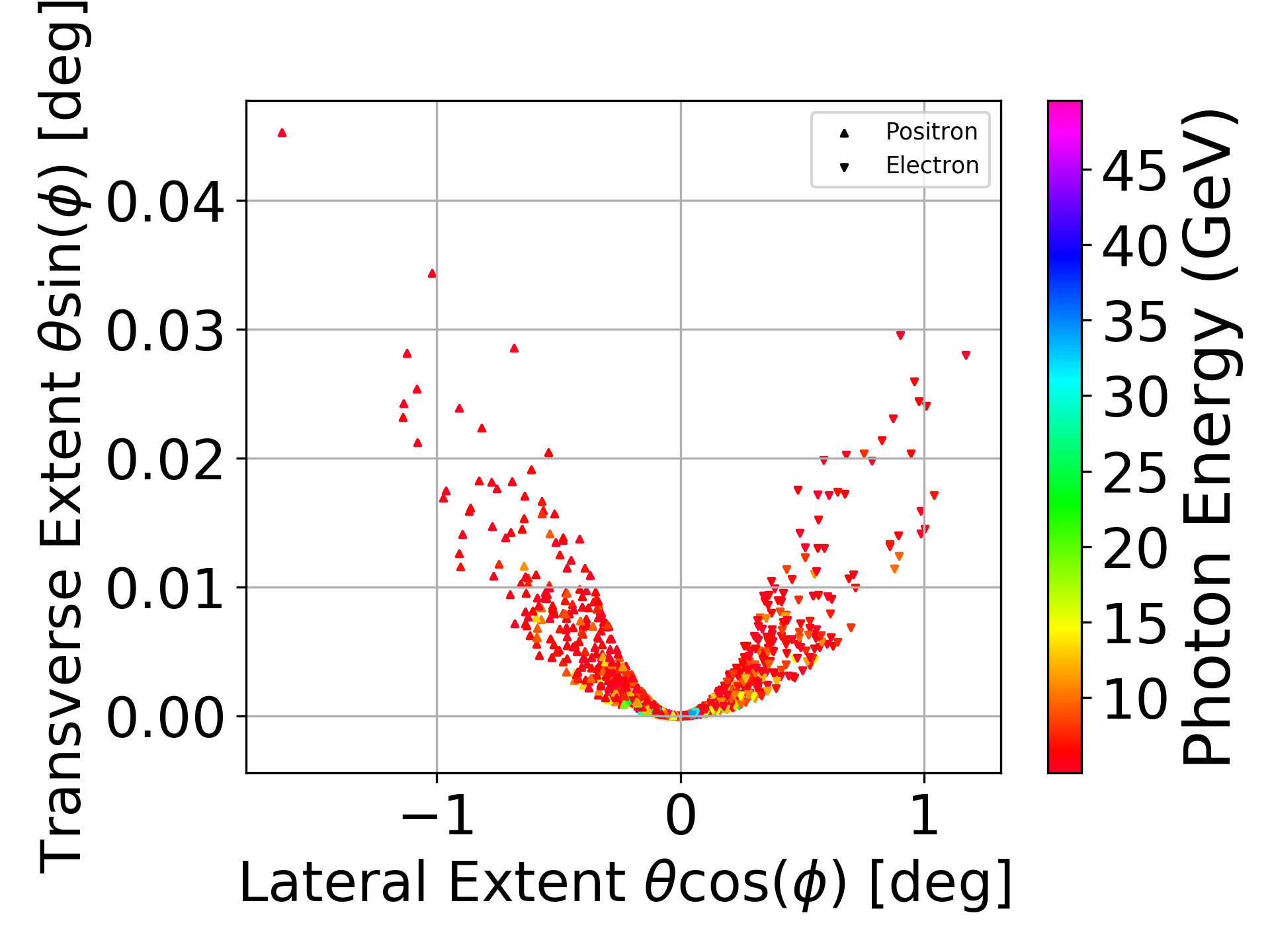}
  	\includegraphics[width=0.45\textwidth]{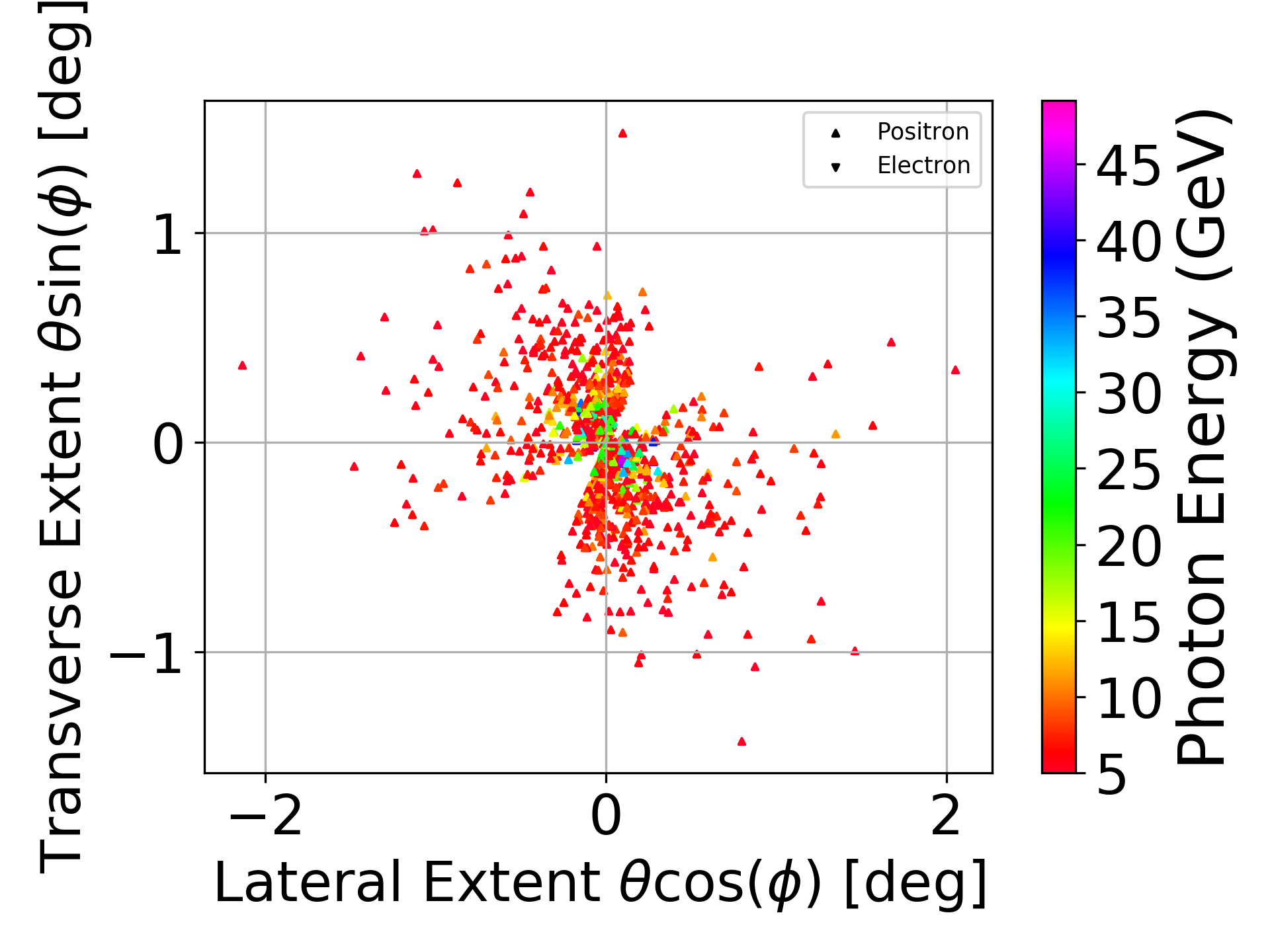}
  	\caption{ Example of halos from blazars in a uniform (left) and maxially helical (right) inter-galactic
	magnetic field as given in Eqs.~(\ref{constantB}) and (\ref{helB}). The colors 
	denote the energy of the observed gamma ray.}
	\label{haloexample}
\end{figure}

\begin{figure}[tbhp]
  	\centering
  	\includegraphics[width=0.45\textwidth]{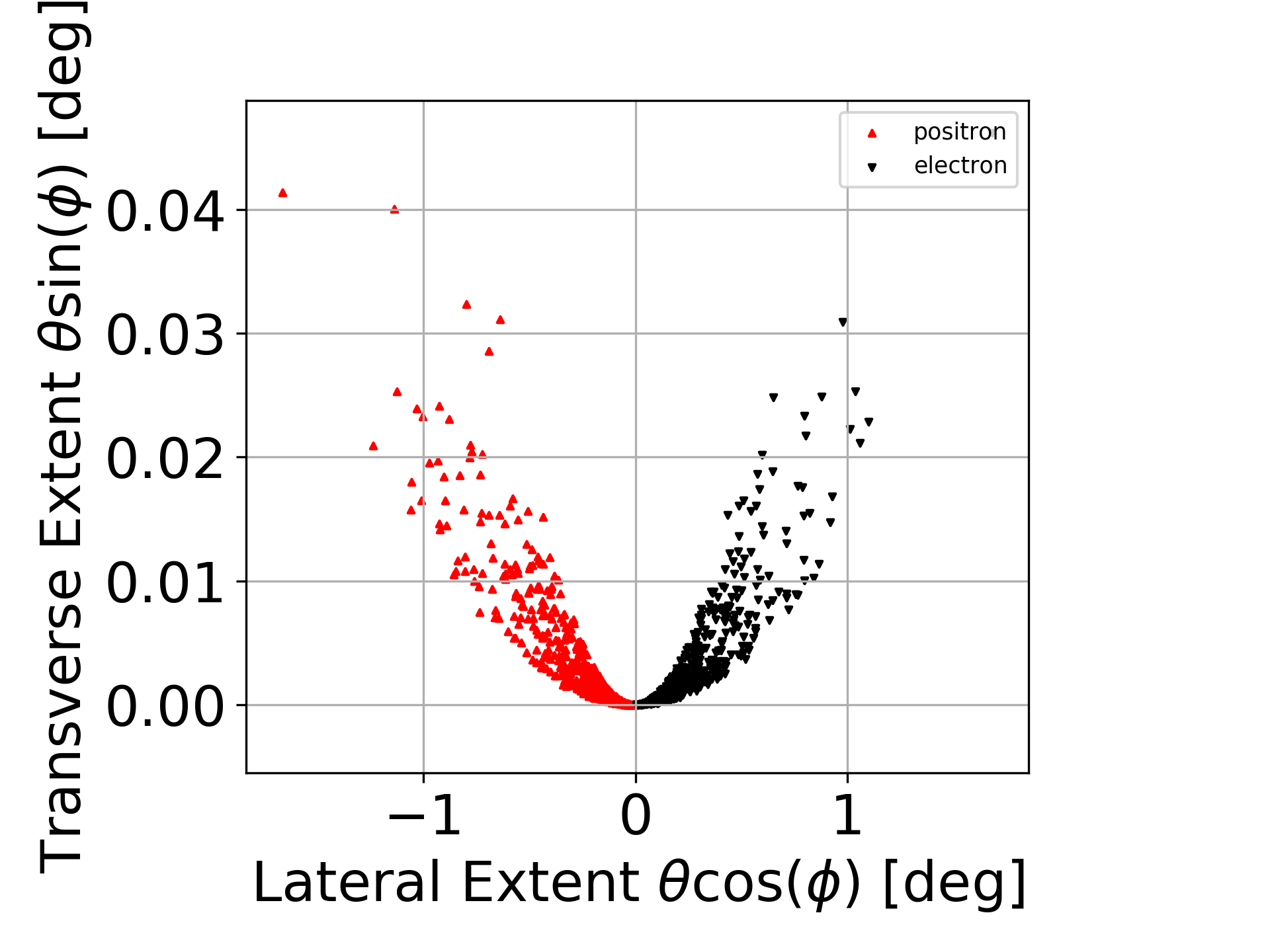}
  	\includegraphics[width=0.45\textwidth]{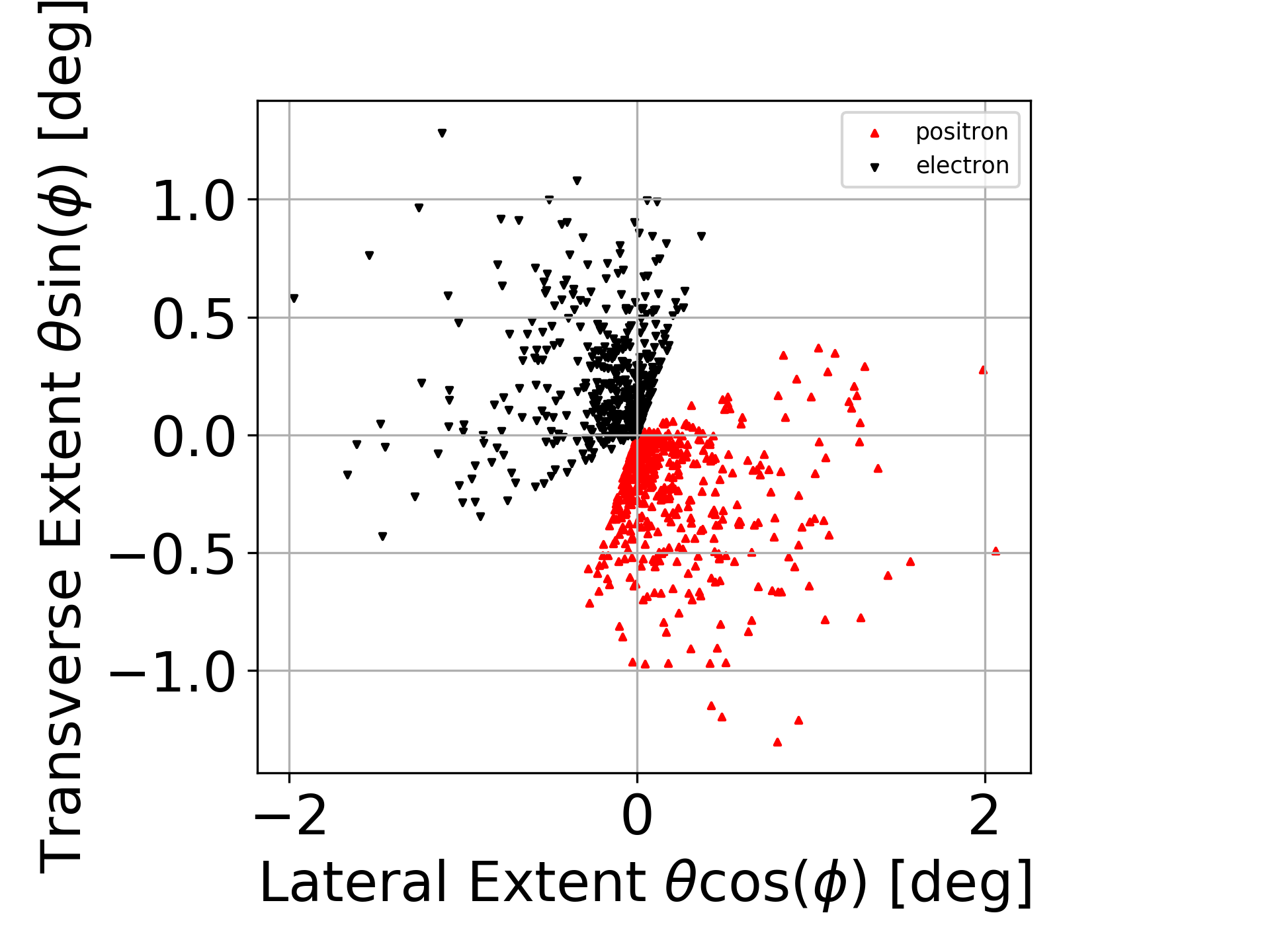}
  	\caption{Same as in Fig.~\ref{haloexample} but now red (black) points originate from inverse Compton
	scattering due to positrons (electrons). If the source was taken to emit along a jet, most of the observed
	gamma rays would originate from either positron or electron processes but not both.}
	\label{haloexample2}
\end{figure}

We show halos for the simple field configurations of Eqs.(\ref{constantB}) and (\ref{helB}) 
in Fig.~\ref{haloexample}. Looking closely at Fig.~\ref{haloexample}, the drawn points are triangular;
upright triangles are gamma rays that originate from electrons and inverted triangles are
those that originate from positrons. The distinction is made clearer in Fig.~\ref{haloexample2}
where red (black) points originate from positron (electron) processes.
If the source was taken to be a jet, gamma rays predominantly from one of the two leptons
will be observed.

\begin{figure}
	\centering
\includegraphics[width=0.45\textwidth]{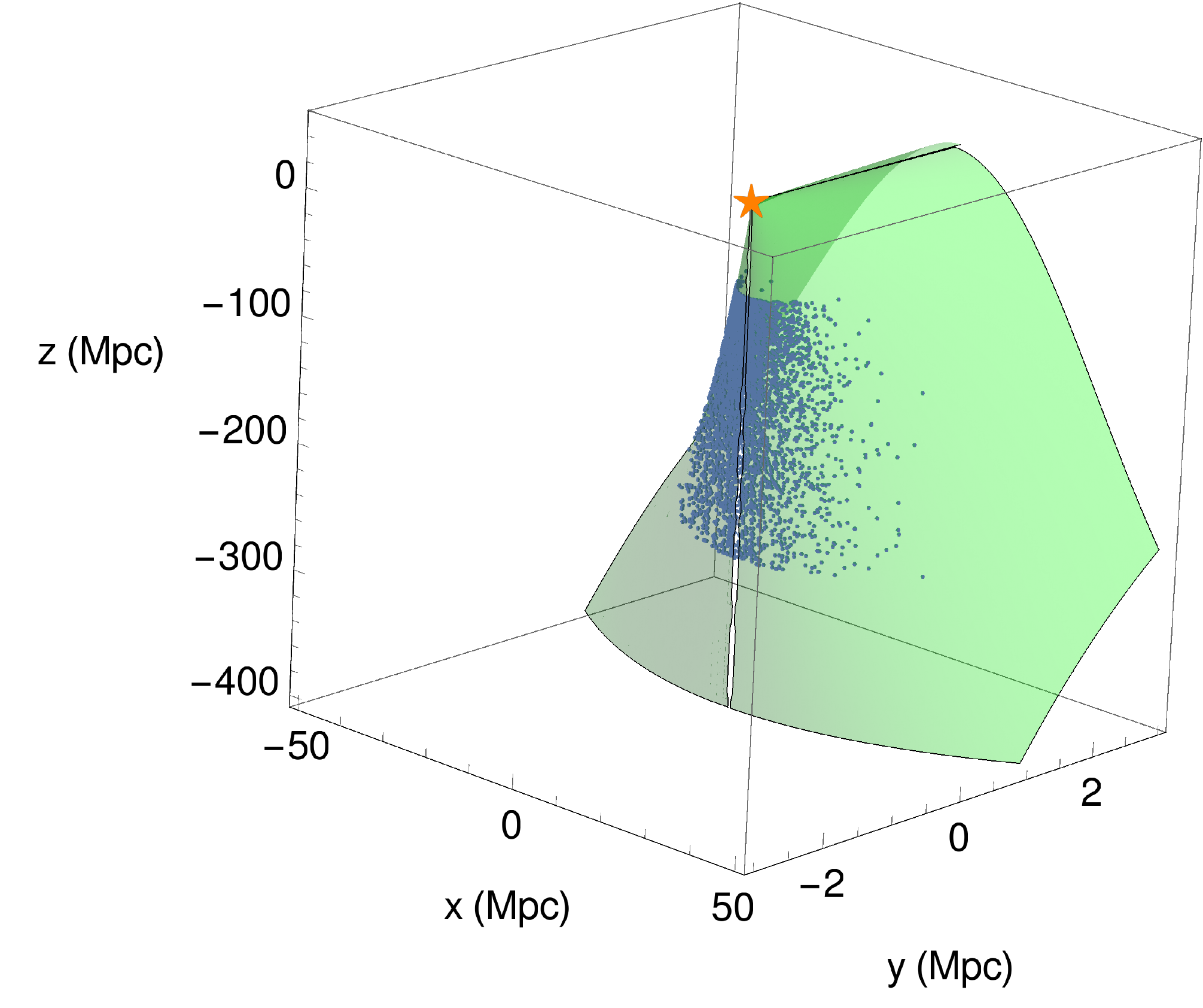}
\includegraphics[width=0.50\textwidth]{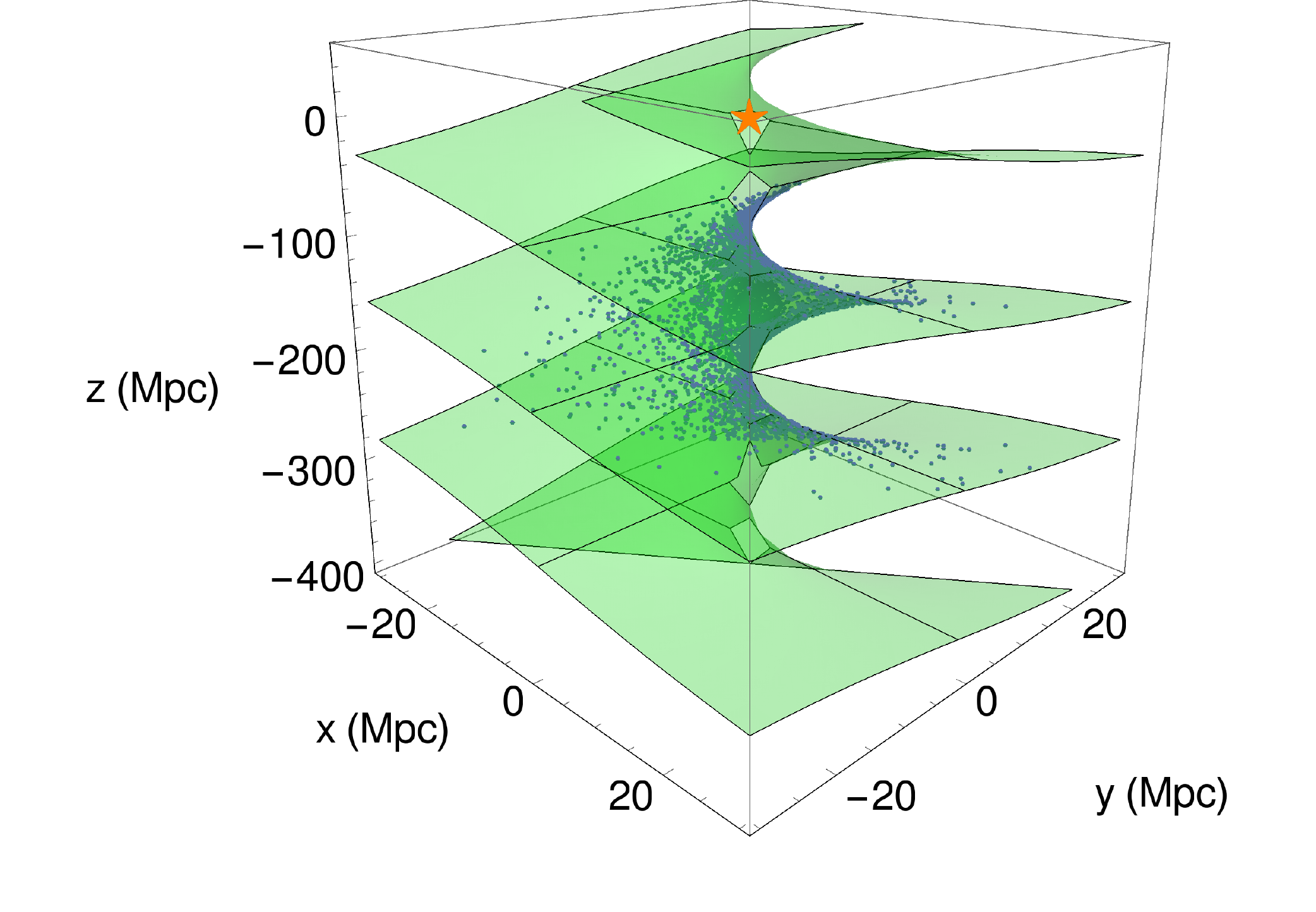}
\caption{
The PP surface for the uniform magnetic field of Eq.~(\ref{constantB}) (left) and
the maximally helical magnetic field of Eq.~(\ref{helB}) (right).
The source is located at the orange star; the observer is at $z=-1~{\rm Gpc}$.
The blue points are the events that give rise to the halos shown in
Figs.~\ref{haloexample}.
}
\label{fluxsources}
\end{figure}

 The constraint equations are quite complicated to solve but there is a helpful visualization.
 First consider the third constraint equation, Eq.~(\ref{const3}), and note that $b_\rho$ and
 $b_\phi$ are also functions of $\theta$, $\delta$ and $\phi$. So Eq.~(\ref{const3}) provides one
 functional relation between these variables that only depends on the magnetic field background.
 Hence the magnetic field defines a two-dimensional surface in space. 
 We will call this the
 ``Pair Production surface'' or the ``PP surface'' since only lepton pair production at this
 surface can send GeV gamma rays to the observer. 
 In Fig.~\ref{fluxsources} we show the PP surface
 for the magnetic fields of Eqs.~(\ref{constantB}) and (\ref{helB}). On these plots we also
 show the pair production locations, ``PP locations'',  that resulted in the halos of Fig.~\ref{haloexample}.
Note that a gamma ray from the source will propagate a certain distance, $d_{\gamma 0}$ and then pair produce. So
 the pair production points also lie on a sphere of radius $d_{\gamma 0}$. 
 This is partly enforced by the law of sines in Eq. (\ref{const1}), which gives a relation between $\delta$ and $\theta$.
 The intersection of this sphere and the PP surface define a one-dimensional curve in space; CMB photons that
 are inverse Compton scattered along the one-dimensional curve can propagate to Earth. 
 However, not all points on this one-dimensional curve will satisfy the final constraint.
 Namely, Eq.~(\ref{const2}), picks out a limited set of points on the one-dimensional curve
 and these give the trajectories of the gamma rays that are observed.

The PP surface can be found analytically for simple cases. For instance, the constraint in 
 Eq.~(\ref{const3}) with the helical magnetic field from Eq.~(\ref{helB}), which has $b_z=0$, reduces to,
 \be
 b_\rho \cos(\delta/2-\theta)=0,
 \ee
 with 
 \be
 b_\rho=\bm{B}\cdot\hat{\bm\rho}=\sin(2\pi z/\lambda+\phi).
 \ee
 As $\cos(\delta/2-\theta)=0$ has only one solution at $\delta=\pi,~\theta=0$
 in the physical range $\theta \in [0,\pi/2]$, $\delta \in [0,\pi]$,
 the surface is mainly determined by $b_\rho=0$ which translates to,
 \be
 \phi=-\frac{2\pi z}{\lambda}.
 \label{analyticPP}
 \ee
 This equation describes a spiral structure as seen in Figure \ref{fluxsources}.

\begin{figure}
	\centering
\includegraphics[width=0.45\textwidth]{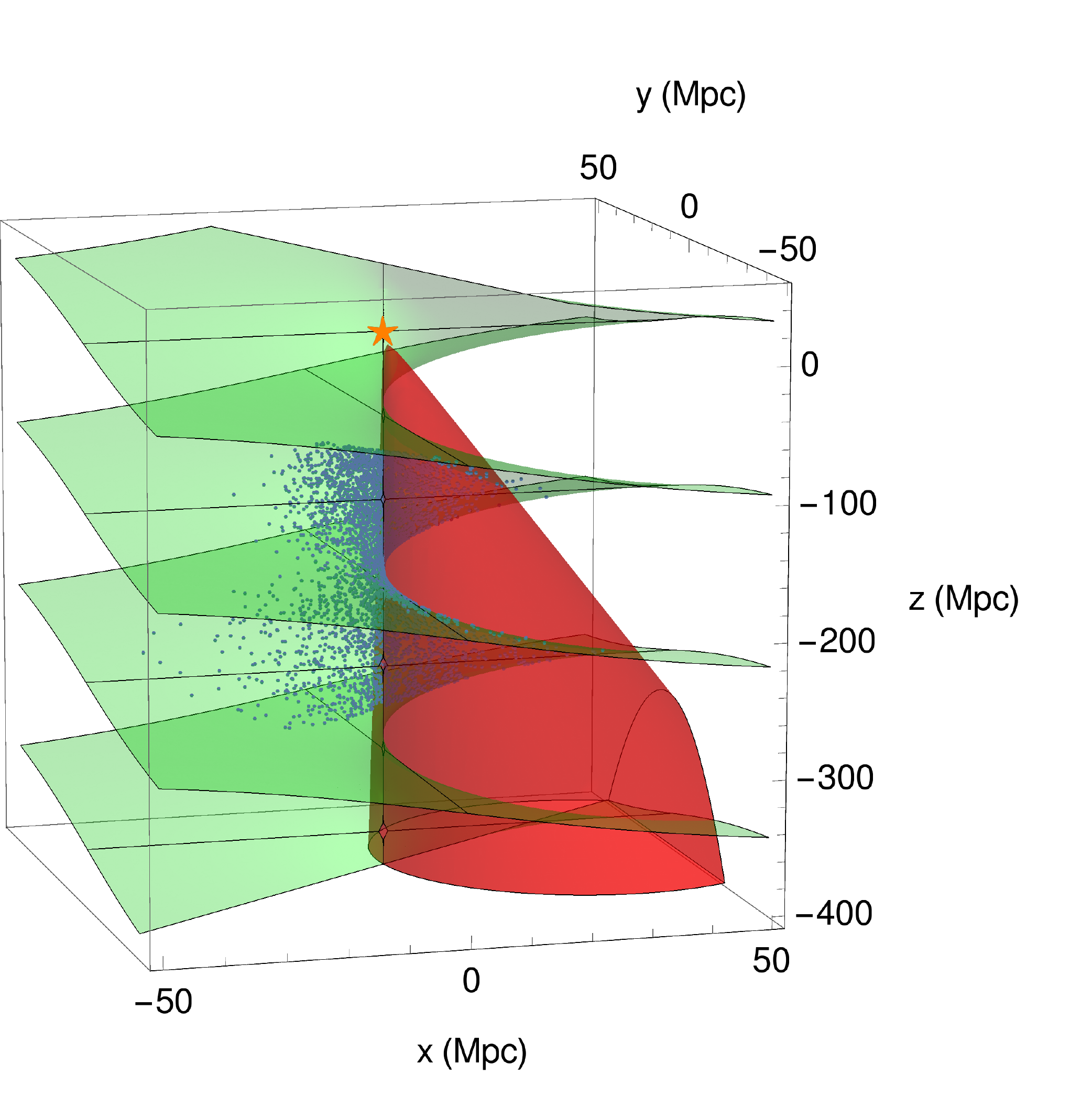}
\includegraphics[width=0.45\textwidth]{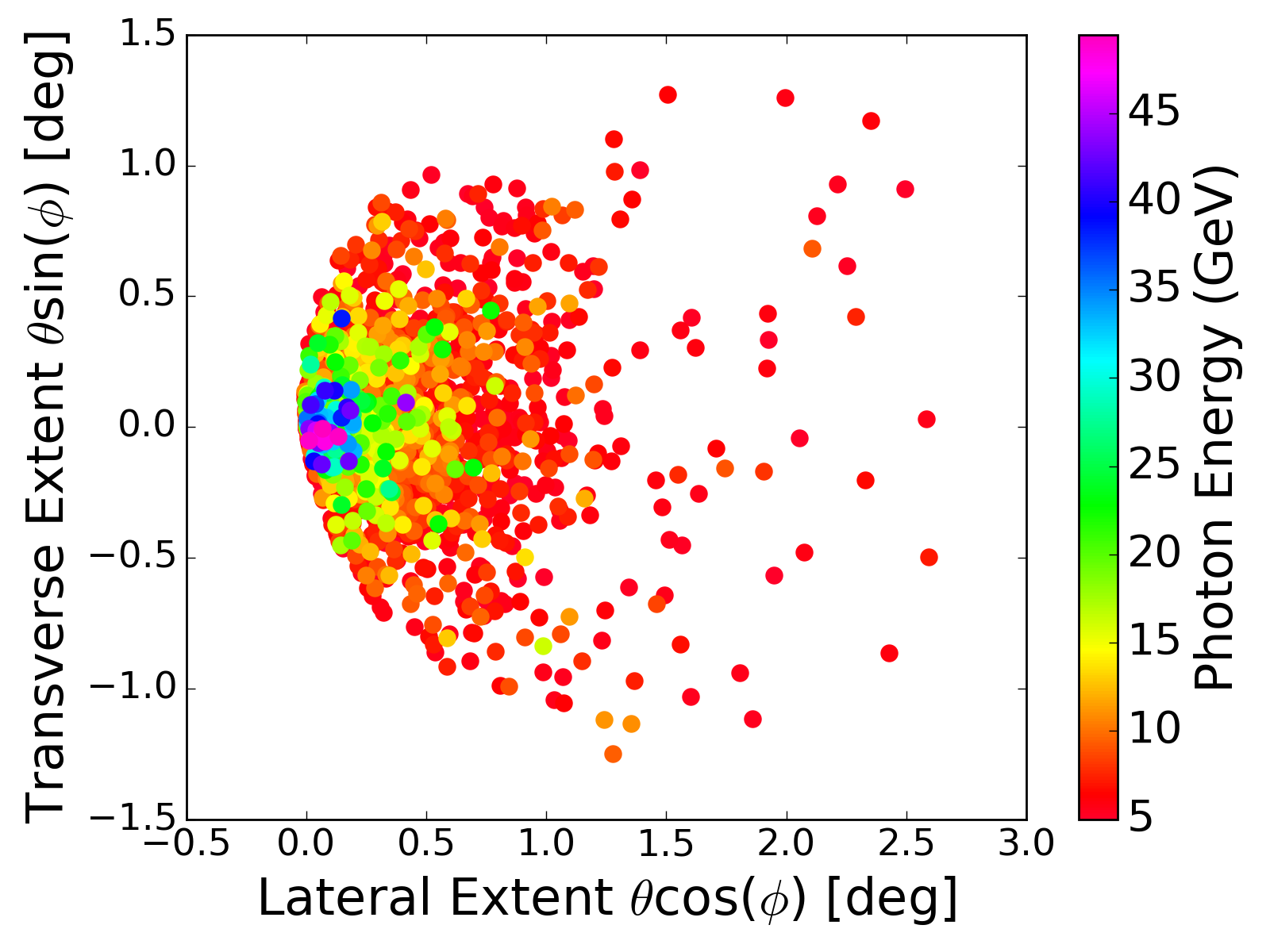}
\caption{Example of how a blazar with a jet will only shine and activate a small region of the 
PP surface (left) and the resultant halo (right). 
The magnetic field is given in Eq.~(\ref{helB}) with $B_0=10^{-14}$G, $\lambda=250\text{ Mpc}$.}
\label{fluxsourcesjet}
\end{figure}

Until now, we have been assuming that the source emits photons isotropically. Below, we will also
 consider the case when the source emits photon in a collimated jet. In that case, there is a fourth constraint 
 restricting the relevant part of the PP surface to where it intersects the jet, and it is quite possible
 that there is no solution. 
 We ignore such cases as they are observationally irrelevant.
 In following figures we will show PP locations, even if they do not lie within the
 jet. Only those PP locations that lie within the jet will lead to observed gamma rays.
For instance, Fig.~\ref{fluxsourcesjet} presents an example in which the source has a 
jet with half-opening angle $\theta_{\rm jet} = 5^\circ$ and the magnetic field is given by Eq.~(\ref{helB}). 
The jet direction is chosen so that the Earth lies within the cone of the jet and the
blazar can be seen directly.
The  left plots in both figures shows that the jet picks out a small region of the PP surface, and the right plots 
show the resulting halo.

\begin{figure}
   		\centering
   		\includegraphics[width=0.45\textwidth]{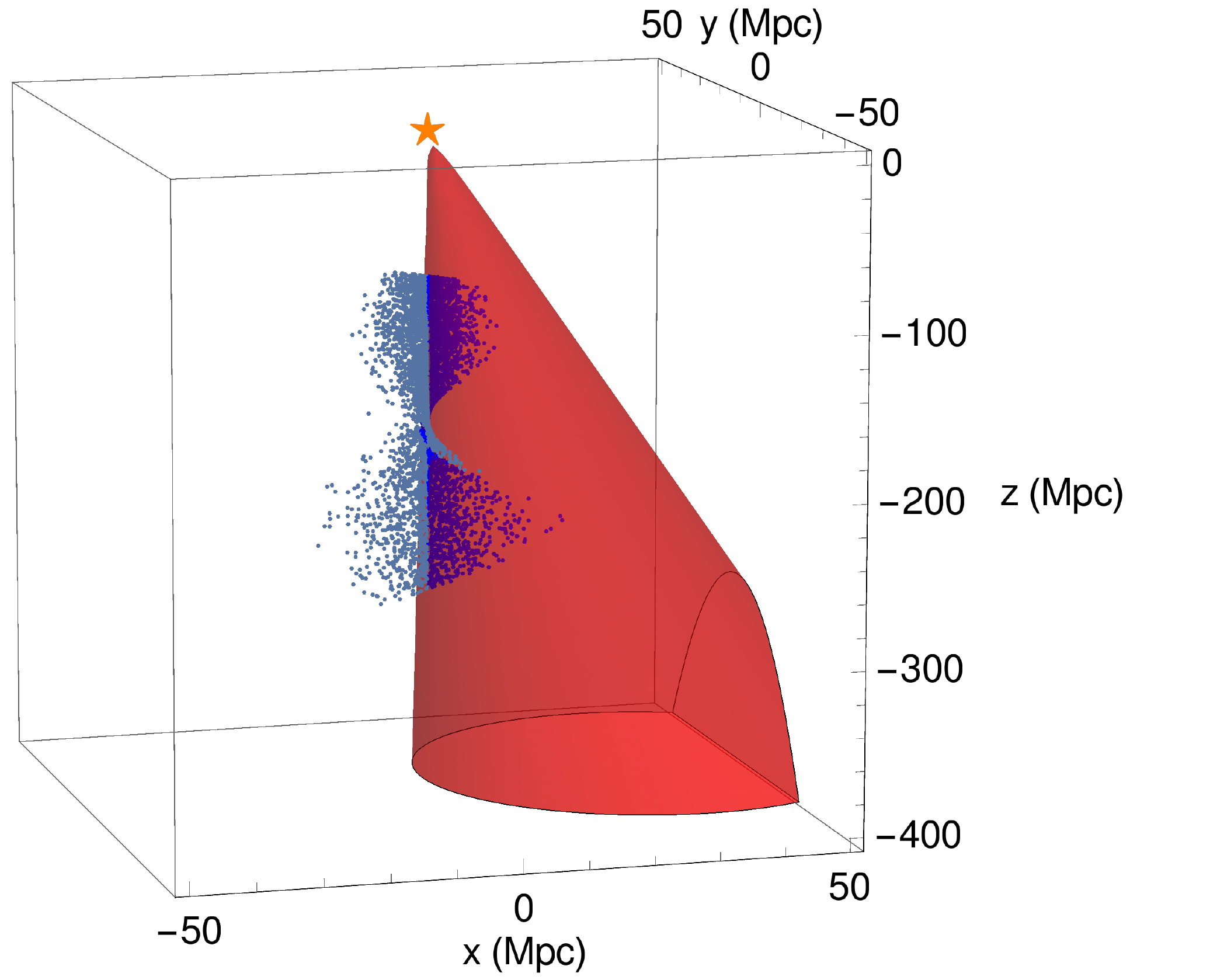}
   		\includegraphics[width=0.45\textwidth]{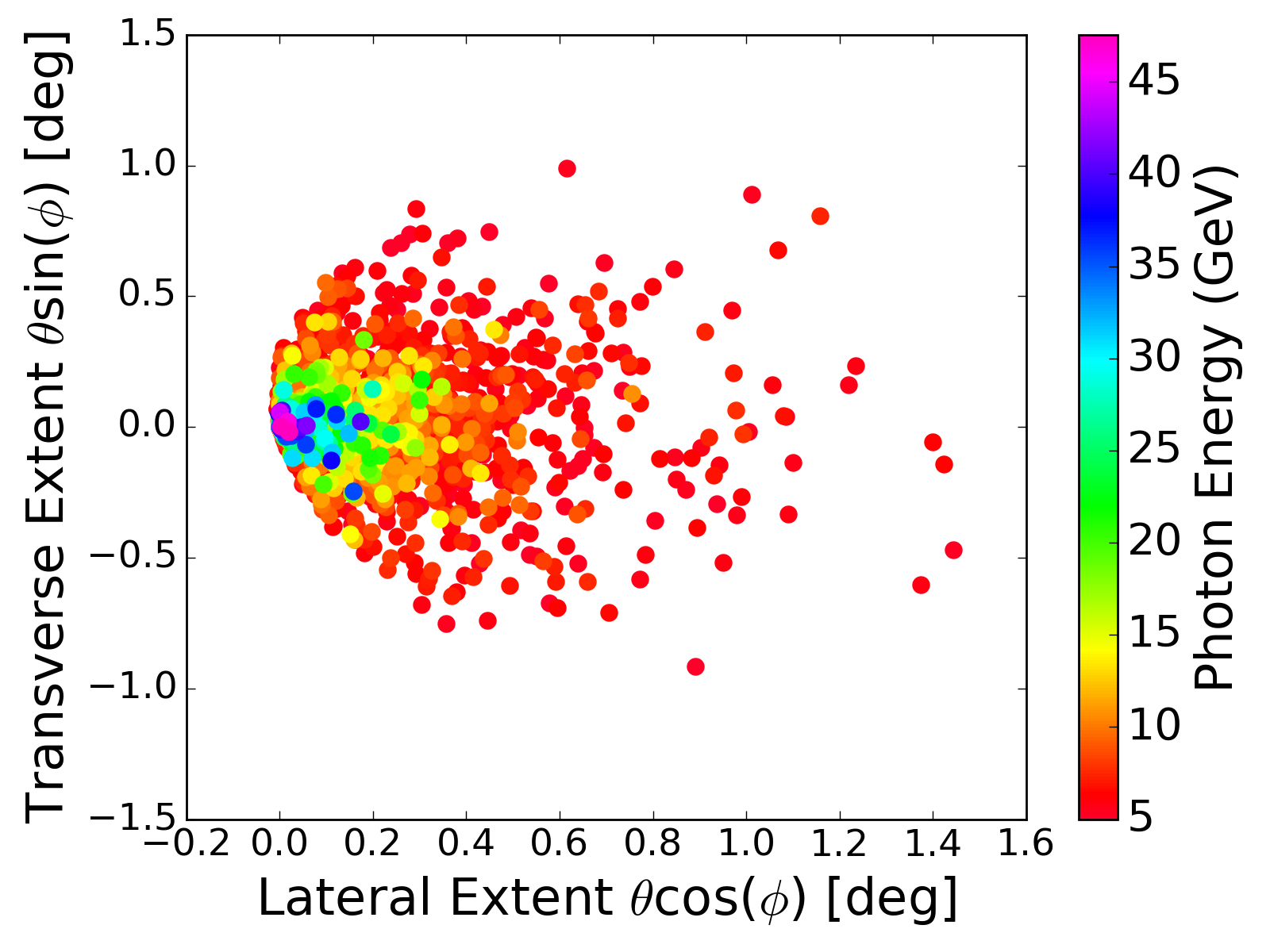}		
   		\caption{
		Monte Carlo simulation with stochastic PDs using the magnetic field of Eq.~(\ref{helB}) and with the
		same setup as in Fig.~\ref{fluxsourcesjet} but with $B_0$ 
		reduced to $5\times 10^{-15}G$. Compared to the right panel of Fig.~\ref{fluxsourcesjet}, 
		we see that the high energy 
		gamma rays (blue and green points) are more clustered and so the halo size is smaller at fixed energy. }
		\label{halo_B_reduced}
		\end{figure}
   	
  \begin{figure*}
  	\centering
  	\includegraphics[width=0.27\textwidth]{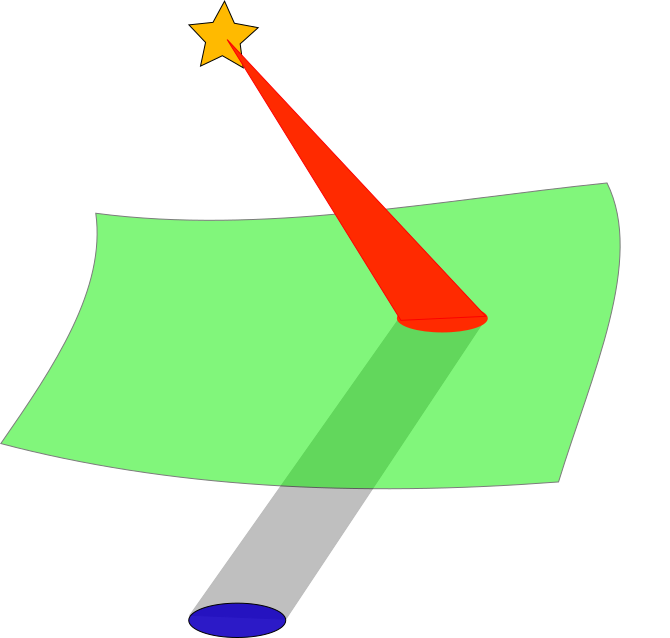}
  	\includegraphics[width=0.27\textwidth]{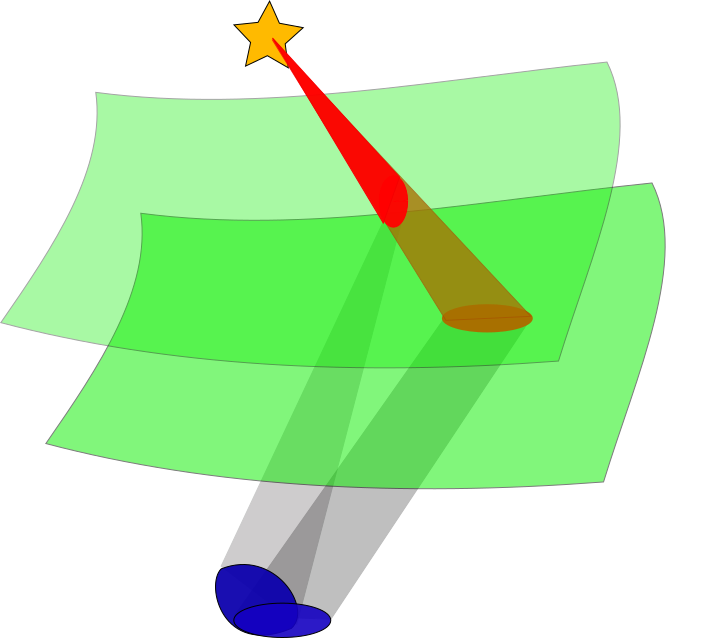}
  	\includegraphics[width=0.20\textwidth]{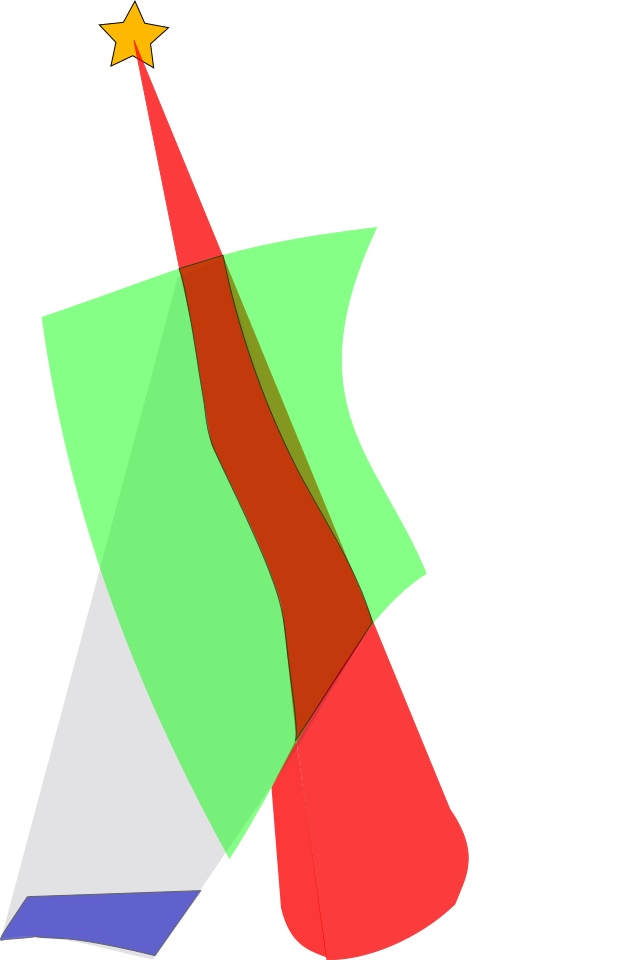}
  	\caption{Sketch of a blazar jet shown in red intersects the green PP surface which delimits 
	the shape of the halo (shown in blue) as seen by some observer. The halo photons must be distributed 
	in the blue region. A situation similar to the one depicted in the middle and third sketch
	can be seen from the simulations in Fig.~\ref{fluxsourcesjet} and Fig.~\ref{varyingcohlength} below. 
	The differences between the many possible shapes arise due the characteristics of the intersection 
	between the blazar's jet and the PP surface.} 
  	\label{haloshapecartoon}
  \end{figure*}

\section{Parameter Dependence of Halo}
\label{halofeatures}	
	
In this section we discuss the structure of the halo as the parameters 
$B_0$, $\lambda$ and the sign of the helicity of the maximally helical magnetic
field in Eq.~(\ref{helB}) are varied. The concept of the PP surface will be a useful
tool for this discussion as it allows us to clearly see how the magnetic field dictates the halo's shape.

The magnetic field strength directly affects the amount of bending of the lepton trajectories
since the gyroradius $R_L \propto 1/B_0$. Therefore a weak magnetic field will require that 
the initial TeV gamma ray is already propagating nearly towards Earth. 
Thus reducing $B_0$ will shrink the size of the halo at any given gamma ray energy,
although lower energy gamma rays may now enter the field of view.
This can be seen in Fig.~\ref{halo_B_reduced} which was created 
using $B_0=5\times 10^{-15}~{\rm G}$ and $\lambda=250~{\rm Mpc}$. The plot looks almost 
identical to Fig.~\ref{fluxsourcesjet}, which
was created using $B_0=10^{-14}~{\rm G}$, except
that the extent of the halo in the $x$ and $y$ directions, for photons of the same energy, has shrunk 
by a factor of $\sim 2$.\\

If one does not track the photon's energy, the effect of a change in $B_0$ is not easily seen through the 
morphology of the halos as their shapes and sizes are determined by the intersection of the jet and 
the PP surface. We show a few examples of this interplay in the sketch of Fig.~\ref{haloshapecartoon}. 
Understanding this could allow us to learn valuable information about the 
inter-galactic magnetic field in the region probed by the PP locations
by observing the halo's shape. 
In a real situation, we cannot observe the full halo
shape as it will be contaminated by background photons coming from
other sources. However, we can still extract certain useful halo information since the 
background is expected to be stochastically isotropic and certainly not parity odd. 

Another important thing to note is that we have assumed the jet and the power spectrum to be 
fixed on the timescale necessary for the creation of the halos. Namely, the path length of two events 
(i.e. the sum of the magnitude of the two vectors shown in Fig.~\ref{halocreation}) will differ as a 
function of their bending angle. Hence this can introduce a significant time delay between the 
subsequent observations of two initial TeV photons emitted from the source at the same time.
If this time delay is large, as would occur for events whose PP locations are Megaparsecs apart,
we would expect the source dynamics to alter its power spectrum and jet direction in that 
timeframe -- making our fixed jet assumption false. 
A quick change in jet direction would translate in observed events arising from potentially very different regions of the PP surface. 
Fortunately, this should not affect our final results once we average over many realizations as this 
already stacks random jet oritentations together. 
A large power spectrum variability could introduce more drastic effects but we will neglect this 
complication in this initial exploration.

Let us quickly comment on the dependence of the morphology on the coherence length of $\mathbf{B}$.
As can be seen from Eq.~(\ref{TeVcoMFP}), the MFP $D^c_{\gamma 0}$ of the TeV photons are of the order of 
$10-100$ Mpc. Any magnetic field with coherence length much larger than $D^c_{\gamma 0}$ will appear constant 
in space. On the other hand, for $\lambda_c \ll D^c_{\gamma 0}$, the halo will be produced from a rapidly varying part of the PP surface
and will be more scattered.

  \begin{figure*}
  	\centering
	        \includegraphics[width=0.35\textwidth]{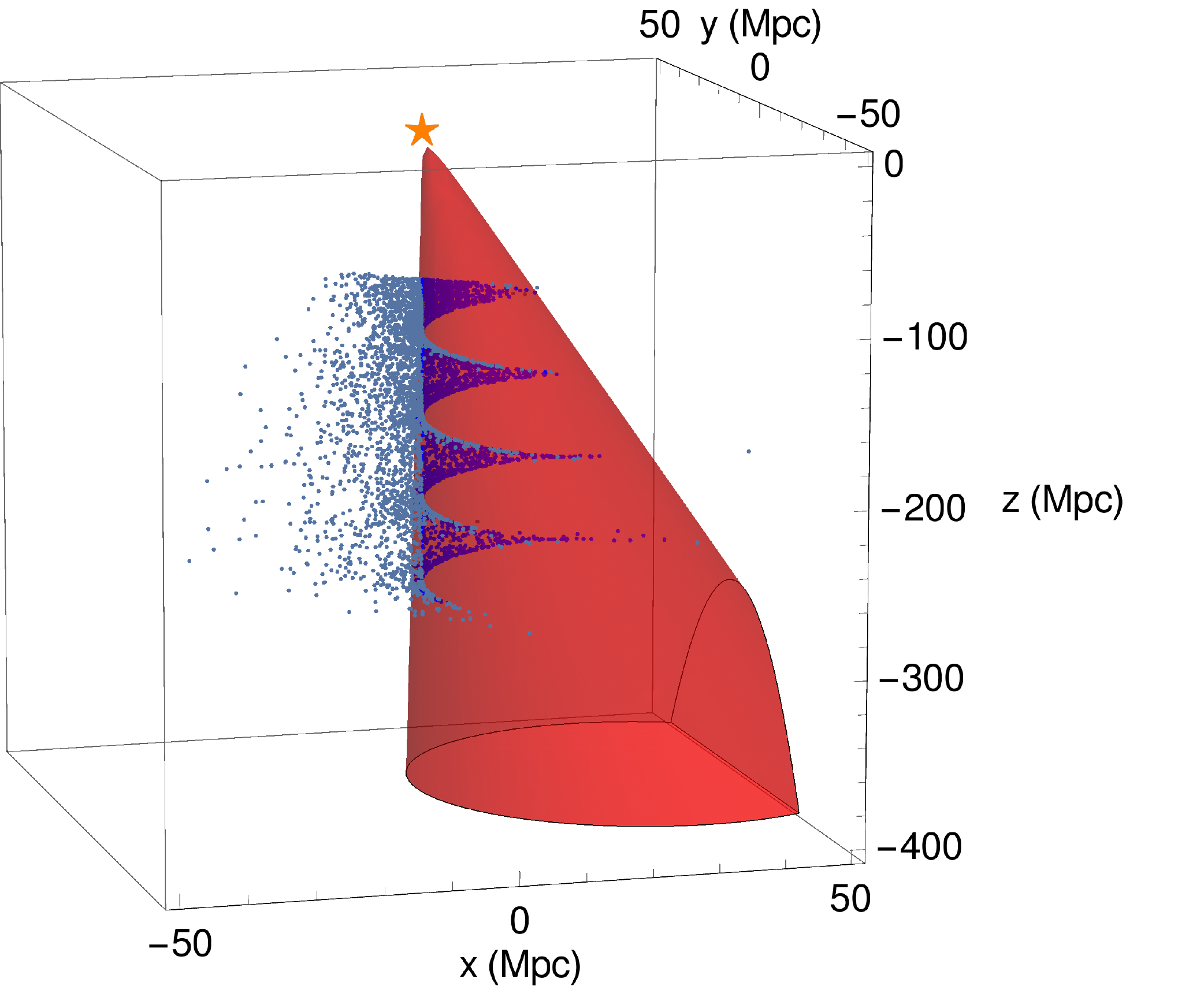}
        	\includegraphics[width=0.35\textwidth]{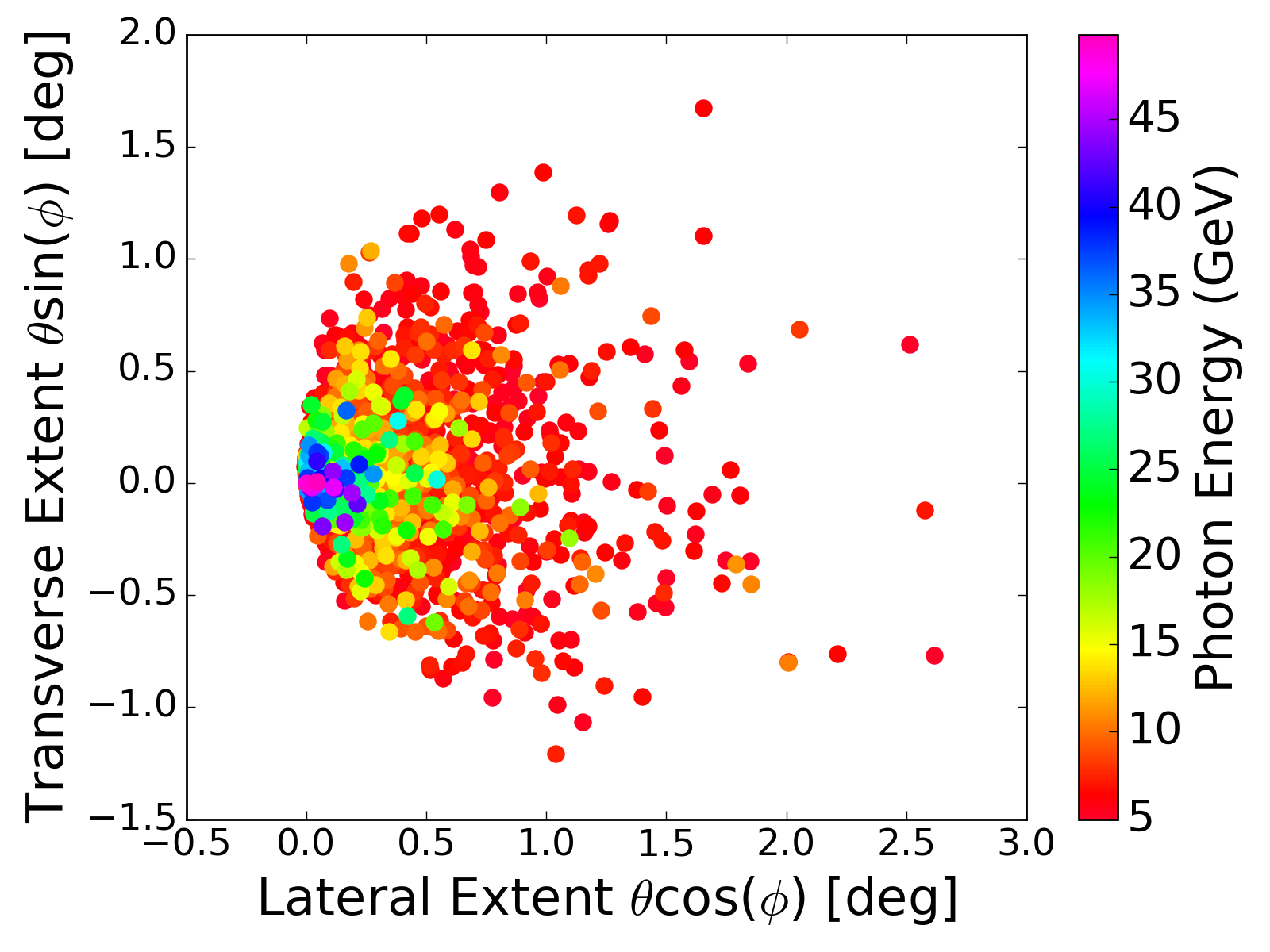}	
        	\includegraphics[width=0.35\textwidth]{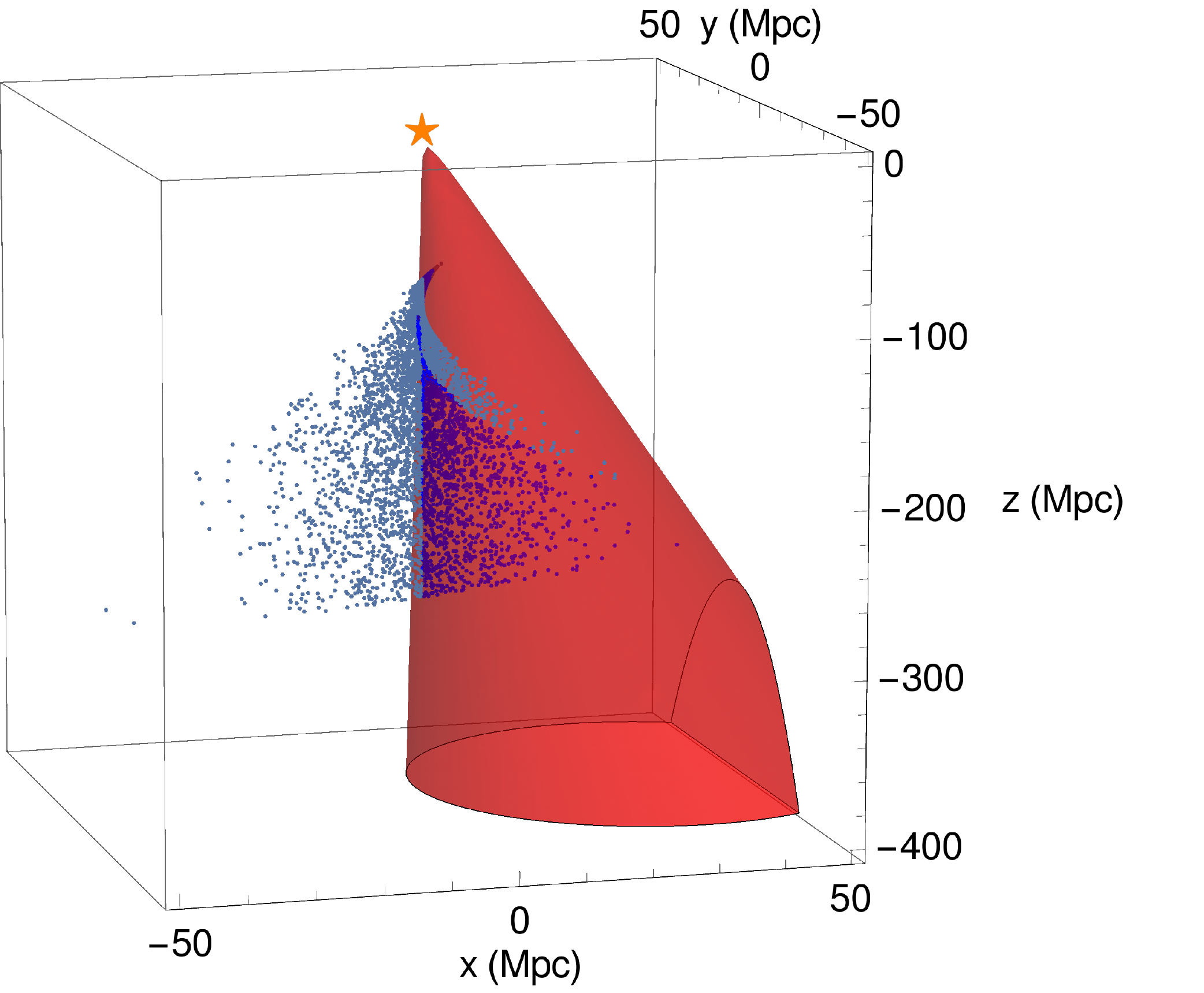}
        	\includegraphics[width=0.35\textwidth]{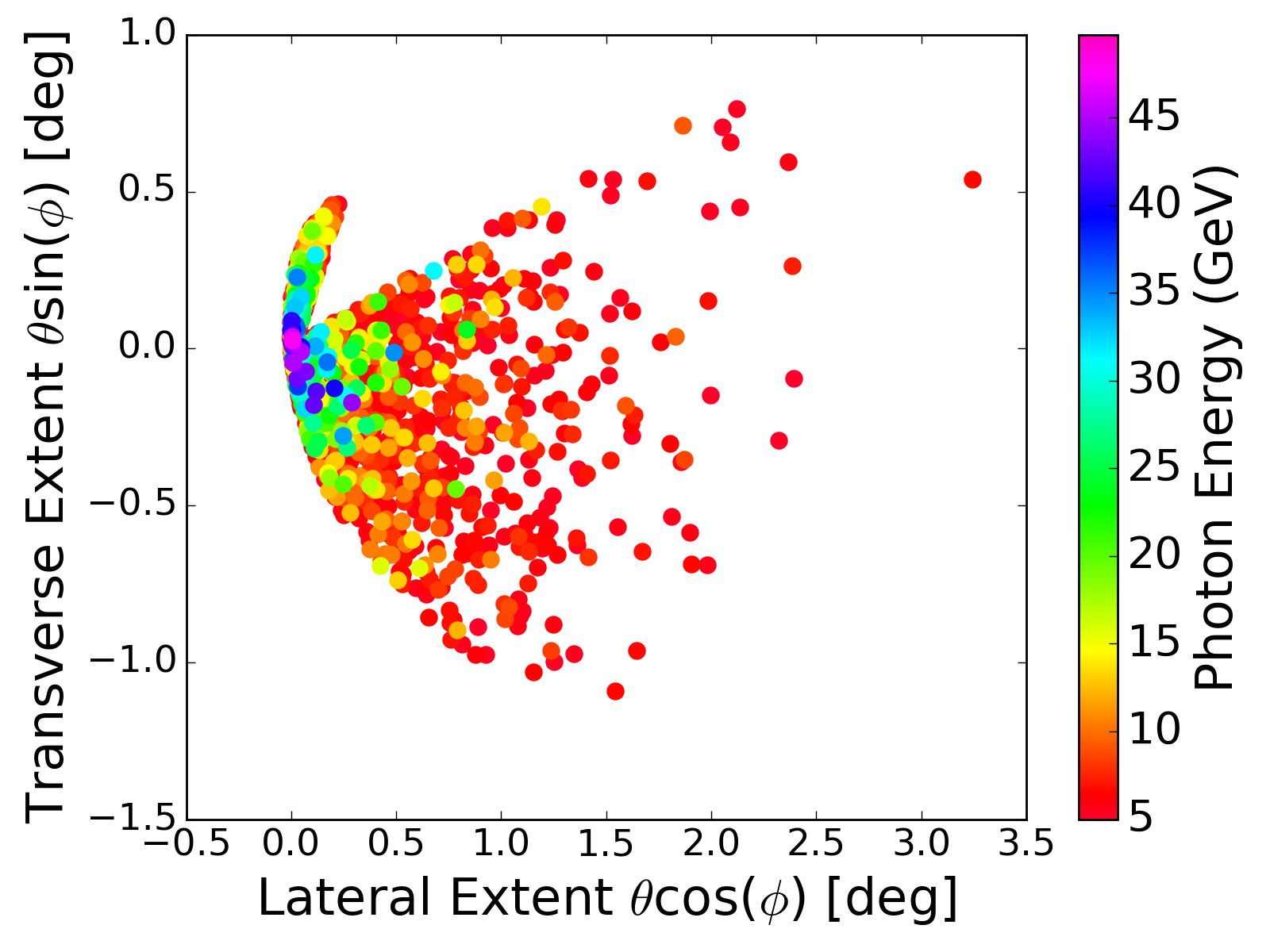}
        	\includegraphics[width=0.35\textwidth]{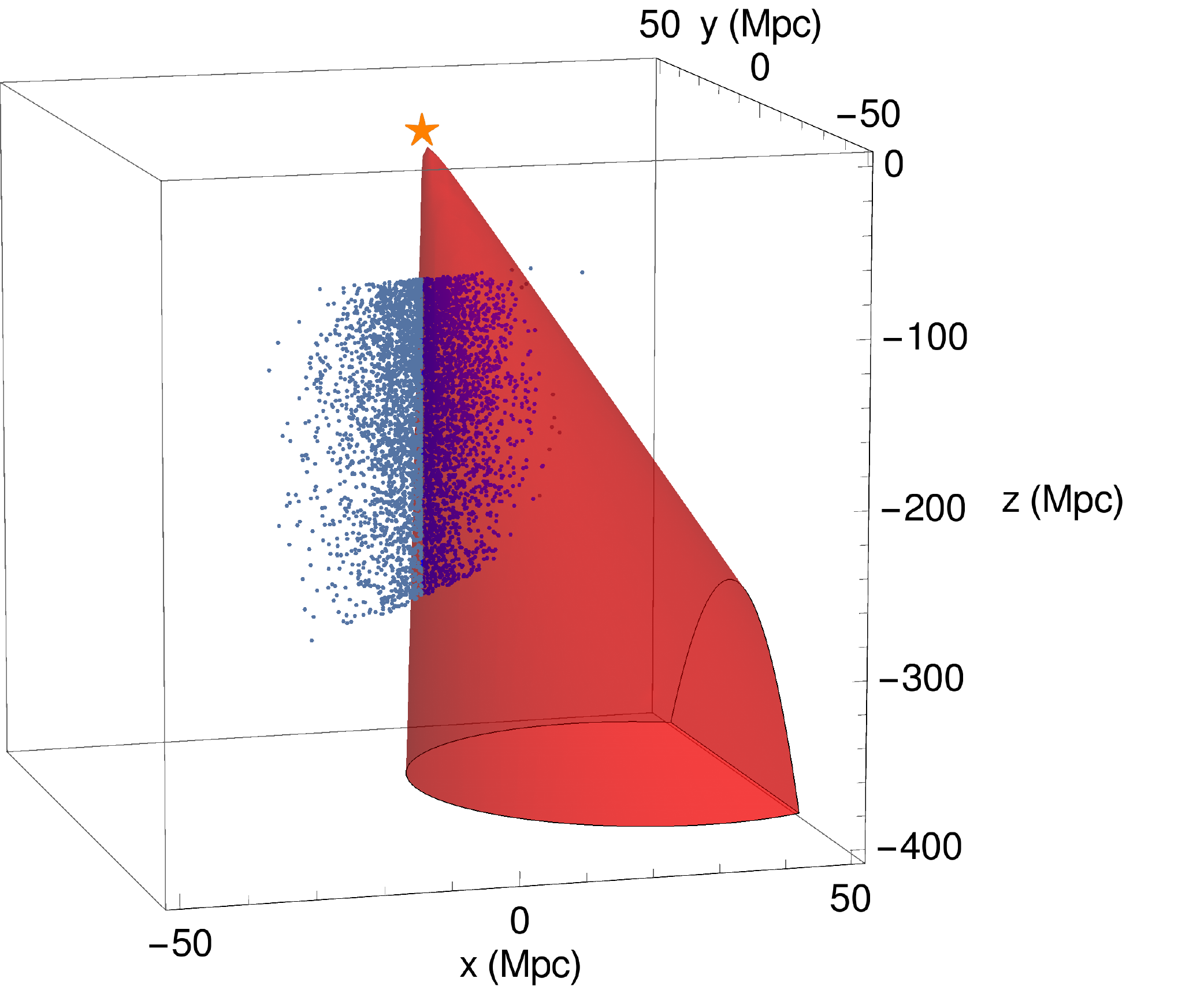}
        	\includegraphics[width=0.35\textwidth]{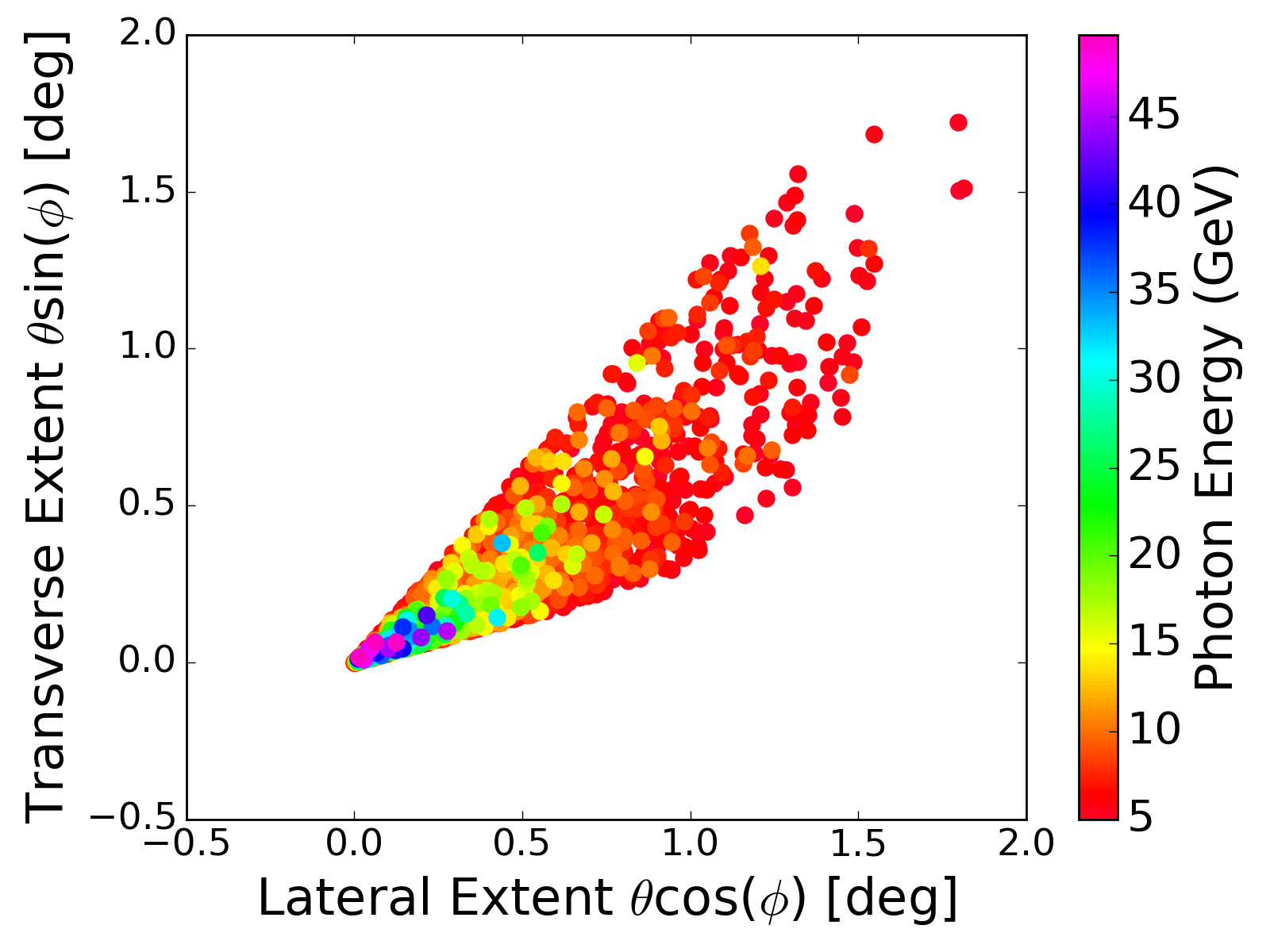}
	        	\caption{The PP locations on the PP surface and jet (left) 
		and corresponding 
		halo (right) for $\lambda=100, 500, 2000~{\rm Mpc}$ for the magnetic field in Eq.~(\ref{helB}) with 
		$B_0=10^{-14}~{\rm G}$. The direction to the source is at $\theta=0$.
		} 
		\label{varyingcohlength}
        \end{figure*}

The helicity of the magnetic field in Eq.~(\ref{helB}) can be flipped by changing $x \to -x$.
A flip in the helicity simply leads to a parity inversion of the PP surface and the halo spiral
also changes handedness. However, to get more statistics, we will investigate both helicities 
using independent simulations in Sec.~\ref{Qstochastic}.
   	
\section{The Q statistic}
\label{Qanalytical}

One of the main goals of this work is to determine if the helicity of the inter-galactic magnetic field can be deduced
from the shape of the blazar halos.
As we have seen, under certain conditions, a helical inter-galactic magnetic field can produce a clear
spiral-like structure in a gamma ray halo. Hence it is important to develop a statistical technique that
is sensitive to this structure. A statistic, called $Q$, was developed in Ref.~\cite{Tashiro:2013bxa}, 
and was applied to the {\it diffuse} gamma ray background observed by the Fermi telescope in 
Refs.~\cite{Tashiro:2013ita,Chen:2014qva}. A non-zero value of $Q$ was observed with 
high confidence in comparison to Monte Carlo simulations that assume no inter-galactic
magnetic field.

One can see from the halo plots above, {\it e.g.} Fig.~\ref{haloexample}, that the arrival direction of 
high energy photons tend to lie closer to the blazar line-of-sight than those for lower energy photons.
Hence different locations of the PP surface are sampled by photons of different energies and 
the observed gamma rays can carry an imprint of any curvature or twist of the PP surface.
More precisely, the work of Ref.~\cite{Tashiro:2013bxa} showed that a left (right) handed helical 
magnetic field will create left (right) handed spiral patterns in the observed photons.

\begin{figure}
\begin{center}
	\centering
\includegraphics[width=0.35\textwidth]{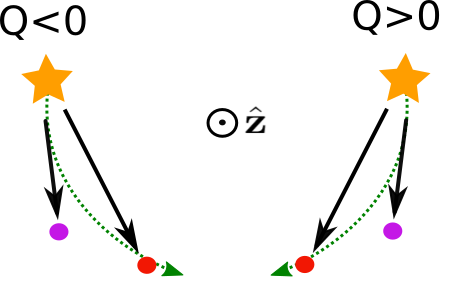} 
\end{center}
\caption{Illustration of the Q-statistic. The plot represents two blazar halos in the observational plane,
each halo with just two photons, one at high energy (purple) and the other at low energy (red).
The image of the blazars is denoted by the stars and the line of sight to the blazar (along ${\hat {\bf z}}$)
points out of the plane of the page as denoted by the arrow tip. The left sketch shows a situation where $Q < 0$ 
since ${\bf n}_{\rm red}\times {\bf n}_{\rm purple}\cdot {\bf n}_{\rm blazar} < 0$; similarly
the sketch of the blazar on the right shows a $Q > 0$ situation.
}
\label{Qidea}
\end{figure}

Below we briefly review the idea behind a slightly modified version of the Q statistic proposed in Ref.~\cite{Tashiro:2013bxa}. We will apply the statistics on regions surrounding an observed blazar whose 
angular position will be denoted by the unit vector $\bf n^{(3)}={\hat {\bf z}}$. We consider a disk of radius 
$R$ centered on the location of the source and consider the set of photons within this disk.
These photons are binned according to their energies into non-overlapping 
bins $\Delta E_1,~\Delta E_2$. We use $N_2$ to denote the number of photons in bin 
$\Delta E_2$ within the disk of radius $R$. We then perform the sum,


\begin{equation}
Q (\Delta E_1,\Delta E_2,R) =
- \mathbf{n}^{(3)} \cdot\Bigg( \frac{1}{N_2}\sum_{j=1}^{N_2}
\mathbf{n}_j^{(2)} \times \Bigg[ \frac{\sum_{i=1}^{N_1} \mathbf{n}_i^{(1)}
 \Theta (\mathbf{m}_i^{(1)}\cdot \mathbf{m}_j^{(2)})}{ \sum_{i=1}^{N_1} \Theta (\mathbf{m}_i^{(1)}\cdot \mathbf{m}_j^{(2)})+\epsilon}\Bigg]\Bigg)
 \label{Qdefn}
 \end{equation}

where $\mathbf{n}_i^{(a)} \equiv \mathbf{n}_i(\Delta E_a)$ is the unit vector denoting the arrival direction 
of photon $i$ in bin $a$; $\mathbf{m}_i^{(a)}$ is the unit vector obtained by projecting 
$\mathbf{n}_i^{(a)}$ on to the xy-plane: $\mathbf{m}_i^{(a)}=\mathbf{n}_i^{(a)}
-(\mathbf{n}_i^{(a)} \cdot \hat{\mathbf{z}}) \hat{\mathbf{z}}$. We've also introduced the infinitesimal quantity $\epsilon$ to keep the denominator from vanishing.
The original Q statistic in Ref.~\cite{Tashiro:2013bxa} 
was defined without the Heaviside function ($\Theta$) in Eq.~(\ref{Qdefn}). 
We illustrate the $Q-$statistic in Fig.~\ref{Qidea}.

\begin{figure}
	\centering
\includegraphics[width=0.40\textwidth]{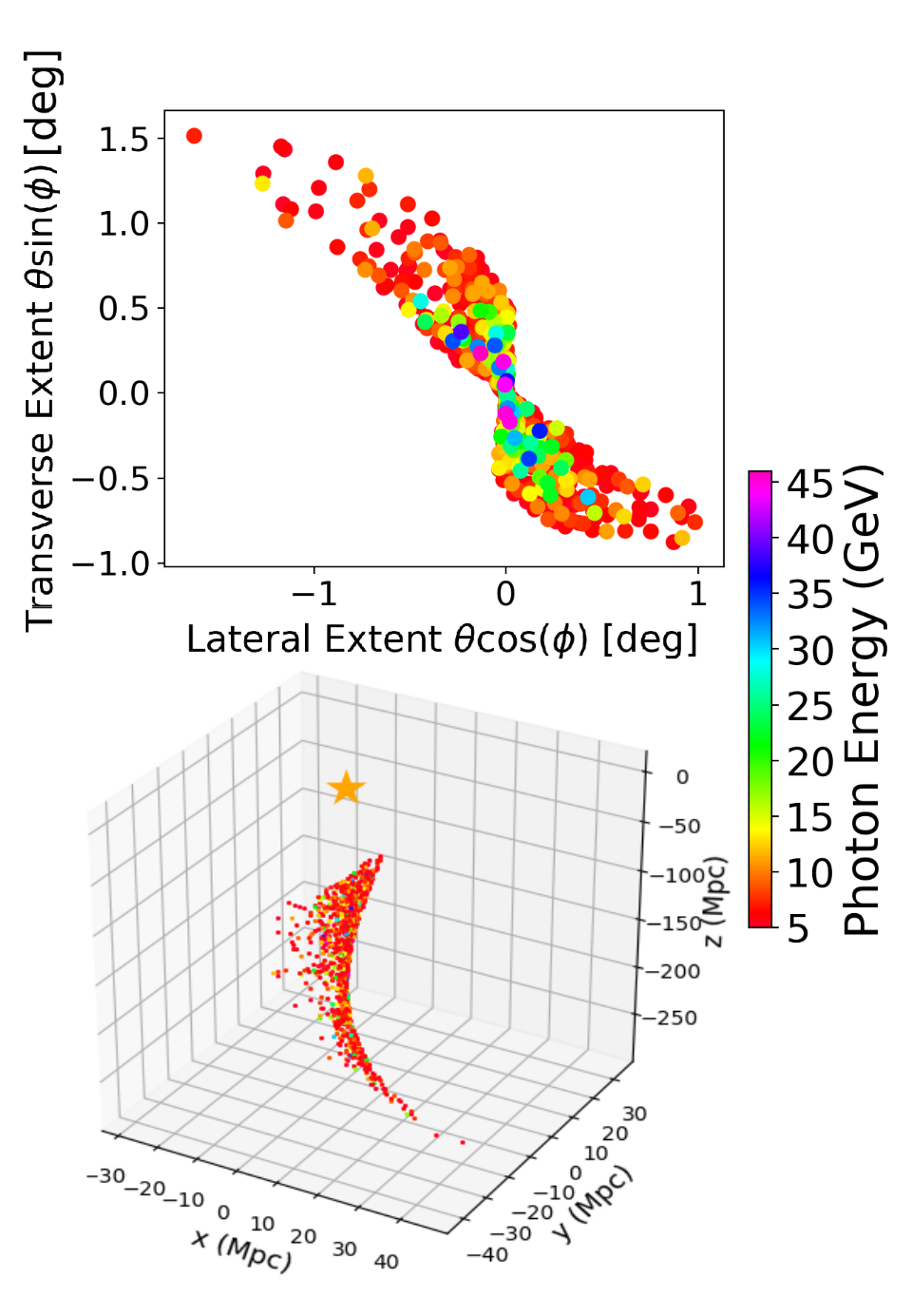} 
\includegraphics[width=0.48\textwidth]{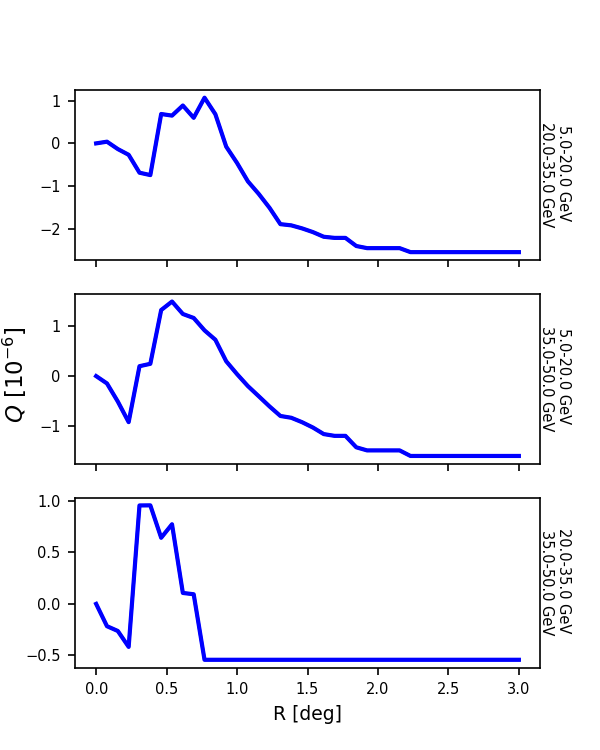} 
\caption{(Top Left) A set of simulated observed photons from the halo formed by a blazar's jet with half-opening angle of $5^\circ$.
The surrounding random magnetic field was created with parameters described in Eq.~(\ref{param1_sim}) and has the form given by Eq.~(\ref{rmfonemode_notappen}). (Bottom Left) The PP locations of the photons responsible for the halo. (Right) The result of applying the Q-statistic to the observed photons on the right.}
\label{nostoQexample}
\end{figure}

In the presence of background gamma rays in addition to the blazar gamma rays,
we expect $Q(R)$ to start near $0$ at $R=0$, grow to a peak value near $R=R_{\text{halo}}$,
where $R_{\text{halo}}$ is the angular radius of the halo, and 
finally come back down towards $0$ at large $R$ where the signal becomes
background dominated. However in mock maps with no background, the value of $Q$ should asymptotically 
flatten out to its maximal value attained at $R_{\text{halo}}$, 
and its value will be negative (positive) for right (left) handed magnetic fields. 
We can see this behavior in Fig.~\ref{nostoQexample} where we simulate the halo without
any stochasticity and with the magnetic field of Eq.~(\ref{helB}).

To showcase the Q-statistic in this paper we will separate the gamma rays in three energy bins:
\begin{equation}
\Delta E_1=(5,20), \ \Delta E_2=(20,35),\ \Delta E_3=(35,50),
\label{energybins}
\end{equation}
all numbers in GeV. This choice was the real reason we only simulated photons between $5$ to $50$~GeV.

\section{The Q statistic Applied to Stochastic Magnetic Fields}
\label{Qstochastic}

The result of Fig.~\ref{nostoQexample} is noisy and can be misleading as we are dealing with random magnetic fields.
 Indeed, these fields can sometimes create halos 
whose Q-statistics suggest the wrong helicity.
It is therefore important to average over many realizations of the magnetic field and the jet orientation. 
Each realization will simulate a blazar with a jet of
half-opening angle $\theta_{\text{jet}}=5^\circ$ and having Earth in its LoS.
The jet is also constrained to generate a halo with at least 3 events in order for the statistics to be applied;
this condition is easily satisfied if Earth is in the jet's LoS. Jets pointing further away from the LoS might still yield observable photons but we would not
be able to identify these blazars and so we don't simulate those cases.

We will consider magnetic fields of the form,   
\be
\label{rmfonemode_notappen}
{\bf B}({\bf {x}})=\frac{1}{2 N^2+2}\sum_{{\bf k}\in K} {\bf b}({\bf k},f_{\text{H}},B_\text{rms})e^{i {\bf k}\cdot {\bf x}}
\ee
with the set $K$ consisting of $2N^2+2$ vectors which have magnitude $k_{\rm mag}$ and whose
directions are approximatively uniformly spread over the unit sphere.
Half of the Fourier coefficients ${\bf b}({\bf k},f_{\text{H}},B_\text{rms})$ are 
drawn from their respective distribution as outlined in Appendix~\ref{Bgeneration}, while the other half
are set by the requirement ${\bf b}({\bf k},f_{\text{H}},B_\text{rms})={\bf b}^*(-{\bf k},f_{\text{H}},B_\text{rms})$,
necessary for obtaining a real value for the magnetic field.
The value of $-1\leq f_{\text{H}}\leq1$ controls the handedness of the field, 
namely $f_{\text{H}}=1~(-1)$ corresponds to a maximally right-handed (left-handed) helical field. Finally 
$B_\text{rms}$ determines the root mean square of ${\bf B}({\bf x})$. 

\begin{figure}
	\centering
	\includegraphics[width=0.45\textwidth]{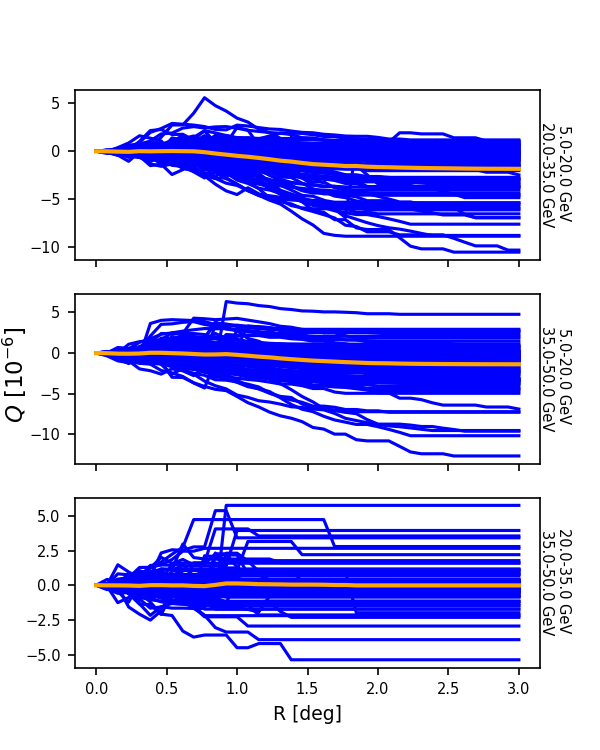}  
	\includegraphics[width=0.45\textwidth]{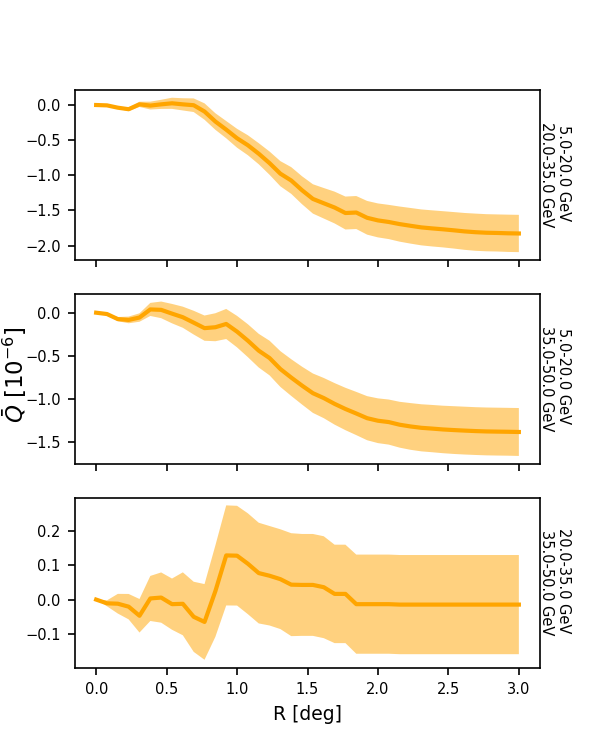} 
	\caption{(Left) $Q(R)$ versus $R$ for 100 Monte Carlo runs when the stochastic magnetic
	fields are generated using the parameters shown in Eq.~(\ref{param1_sim}) with $f_H=+1$. 
	The mean ${\overbar Q}(R)$ is shown by the orange curve.
(Right) A zoomed-in view of ${\overbar Q}(R)$. The width of the error band is 
	given by the standard error {\it i.e.} standard deviation of the 100 Monte Carlo $Q(R)$
	values divided by the square root of the sample size (100).
	}
	\label{nostoQ2}
\end{figure}

In Fig.~\ref{nostoQexample} we compute the Q-statistics for 100
realizations of halos created with a random magnetic field created using the parameters
\ba
\label{param1_sim}
B_{\rm rms}&=&1\times 10^{-14}\, \text{G},\ k_{\rm mag}=0.01/\text{Mpc}, \nonumber \\
f_{\text{H}}&=&+1, \ 2N^2+2=27,
\ea
where $2N^2+2$ is the number of directions of the ${\bf k}$ vector in Eq.~(\ref{rmfonemode_notappen}).
In Fig.~\ref{nostoQ2} we plot the average of $Q$ over all the realizations,
denoted ${\overbar Q}(R)$, for the same runs as in Fig.~\ref{nostoQexample} and for the three 
gamma ray energy bin combinations.
By doing so, we have in mind of averaging the Q-statistics obtained from small regions around multiple 
observed blazars.
The plot also shows the standard error in ${\overbar Q}(R)$
which is given by the standard deviation of $Q(R)$ divided by the square root of the number
of realizations in the Monte Carlo simulations.
The standard error follows from the central limit theorem and is the error in using the sample
mean to estimate the population mean. 
However, it assumes that the samples, Monte Carlo simulations in our case, from which values of
$Q(R)$ are drawn are independent and identically distributed. This is certainly true in our setup but
may not be true for actual observations in which 
the same photons might contribute to the $Q(R)$
calculated for blazars that are close to each other. In addition, there will be variation in  
the distance to observed blazars and other source characteristics. We plan to take some of these
factors into account in a follow-up analysis.

\begin{figure}
	\centering
	\includegraphics[width=0.48\textwidth]{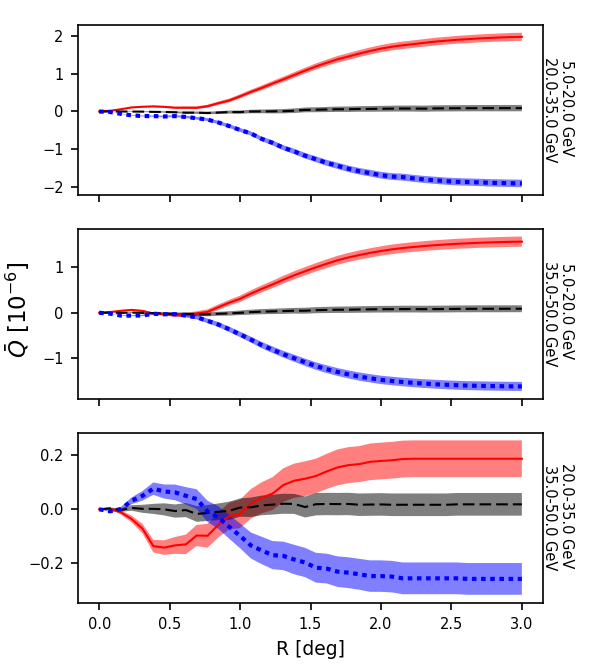} 
	\includegraphics[width=0.48\textwidth]{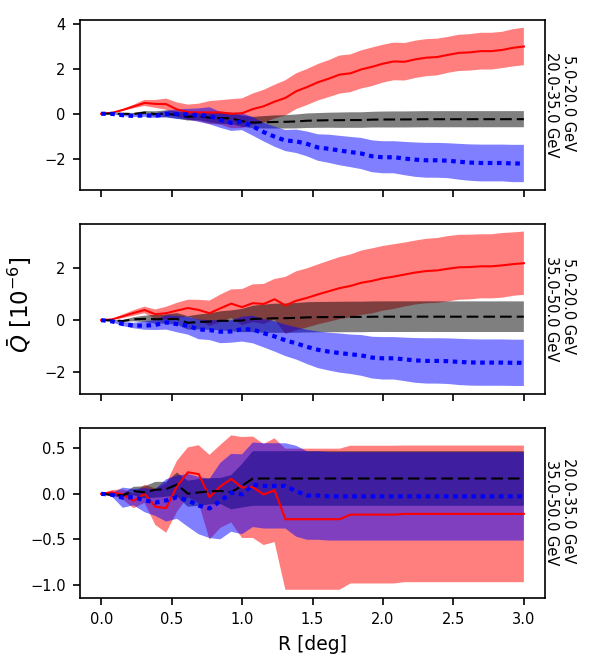} 
	\caption{${\overbar Q}(R)$ versus $R$ when averaging for 1000 simulations (left) and 20 simulations (right). The
realizations had parameters $f_{\text{H}}=-1$ (red, solid), $f_{\text{H}}=0$ (black, dashed)
	and $f_{\text{H}}=+1$ (blue, dotted) for the three energy bin combinations: $(\Delta E_1,\Delta E_2)$ (top row),
	$(\Delta E_1,\Delta E_3)$ (middle row) and $(\Delta E_2,\Delta E_3)$ (bottom row) where the bins are
	defined in Eq.~(\ref{energybins}). The other parameters of the stochastic magnetic fields 
	are given in Eq.~(\ref{param1_sim}) and the bands denote standard error of ${\overbar Q}$.
	}
\label{uniB_Q_100_hel1}
\end{figure}

\begin{figure}
	\centering
	\includegraphics[width=0.48\textwidth]{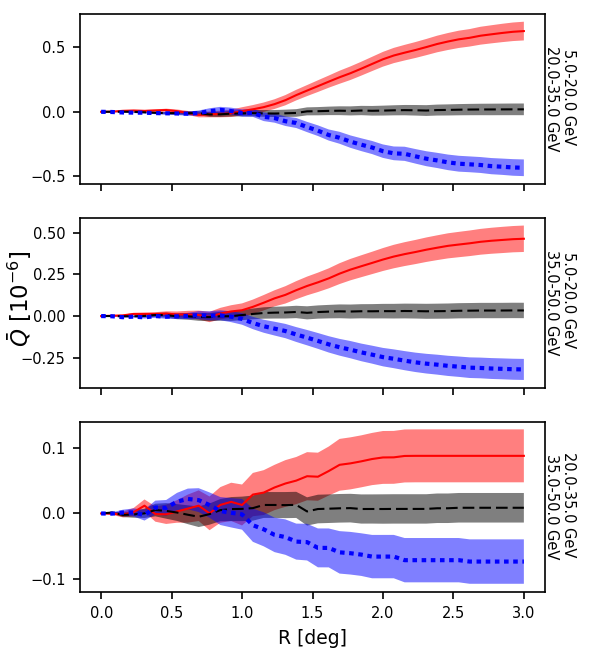} 
	\includegraphics[width=0.48\textwidth]{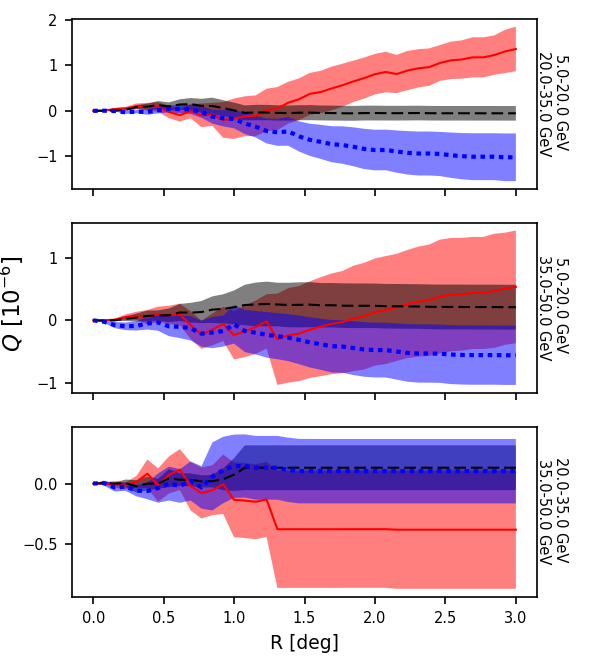} 
	\caption{Analysis of the same data as used for Fig.~\ref{uniB_Q_100_hel1} but computed with the original definition of $Q$. Namely, the 
		$\Theta$ term in Eq.~(\ref{Qdefn}) is simply replaced by the value 1. It is clear both from the magnitude and
		the size of the error bars that the modified $Q$ is a sharper statistics.
	}
\label{oldQ}
\end{figure}

Pushing the statistics further, we are clearly able to distinguish between many different properties of the magnetic field as the number of realization increases. For instance, we have plotted ${\overbar Q}(R)$ 
versus $R$ for 1000 Monte Carlo simulations for $f_H=0,\pm 1$ (Fig.~\ref{uniB_Q_100_hel1}) and
the plots show a clear correlation between ${\overbar Q(R)}$ and the helicity of the magnetic field. We can also
notice distinct oscillations that occur at small $R$.
It is also reassuring that the $f_H=+1$ plot is the mirror image of the $f_H=-1$ curve, just as we would
expect due to parity reflection. Within the clean context studied here, we only need $\sim 10-20$ halos before 
we can detect the sign of the helicity through the sign of the $Q$'s at large $R$.
However this number depends heavily on the properties of the magnetic field, source variability and background. Hence the determination of the exact amount of data 
required to make such detection will require a careful analysis of these parameters and therefore is relegated to future work.

\begin{figure}
	\centering
	\includegraphics[width=0.5\textwidth]{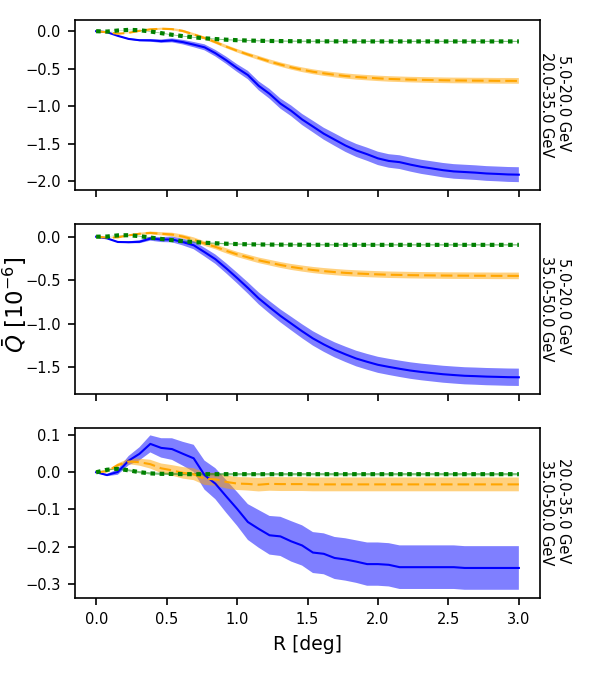} 
	\caption{${\overbar Q}$ versus $R$ for
	$1000$ simulations with $B_{\rm rms}=10^{-14}~{\rm G}$ (blue,solid), 
	$5\times 10^{-15}~{\rm G}$ (orange, dashed), and
	 $2\times 5\times 10^{-15}~{\rm G}$ (green, dotted) and $f_H=1$, $k=0.01/{\rm Mpc}$.
	 }
\label{Q_Bvar_k0d01_fh1_nolegend}
\end{figure}

In Fig.~\ref{oldQ}, we show that the $\Theta$ factor in our definition of $Q$ in Eq.~(\ref{Qdefn}) 
improves the resolution.
Without the $\Theta$ factor, the Q-statistic is determined by the cross product of the {\it average} 
arrival direction of the photons (in two energy bins) within a radius $R$ of the source. 
Because of the electron-positron symmetry, photons tend to arrive on either side of the
source (see Fig.~\ref{haloexample2}), and the average arrival direction tends to be near the origin.
Introducing the $\Theta$ factor ensures that for every high energy photon selected, 
we only average the low energy photons that arrive on the same side with respect 
to the source. Then there is a larger contribution to the value of $Q$.
Essentially the $\Theta$ term limits the sum to gamma rays within the 
electron (or the positron) branch of the halo (see Fig.~\ref{haloexample2}).

Next we examine the dependence of ${\overbar Q}(R)$ on magnetic field parameters.
In Fig.~\ref{Q_Bvar_k0d01_fh1_nolegend} we plot ${\overbar Q}(R)$ for several different
magnetic field strengths and for fixed helicity $f_H=1$. We see that increasing the magnetic
field strength leads to an increasing amplitude of ${\overbar Q}$ for {\it all} energy combinations. The 
increase is due to the magnitude of $\bf{n}_1\times \bf{n}_2$ which becomes larger as the bending 
allows $\bf{n}_1$, $\bf{n}_2$ to point further apart.

\begin{figure}
	\centering
	\includegraphics[width=0.5\textwidth]{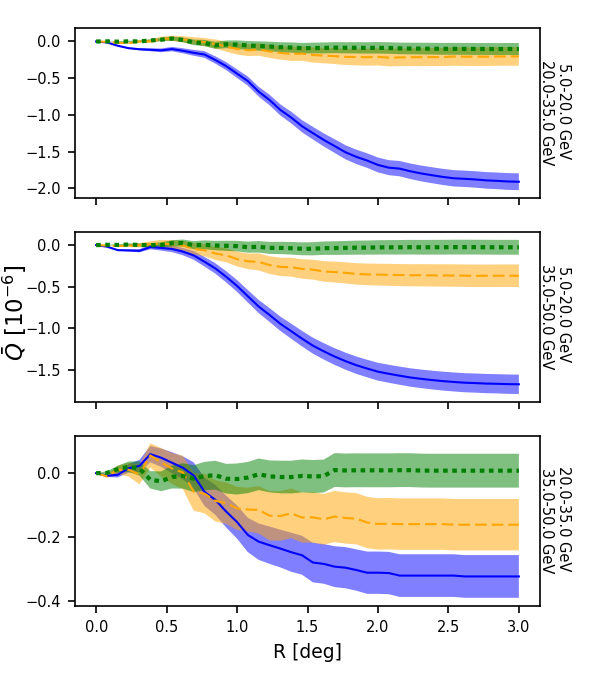} 
	\caption{${\overbar Q}$ versus $R$ for $1000$ simulations with 
	$k=0.01/{\rm Mpc}$ (solid, blue), $k=0.05/{\rm Mpc}$ (orange, dashed), $k=0.1/{\rm Mpc}$ (green, dotted)  
	with $B_{\rm rms}=1\times 10^{-14}~{\rm G}$ and $f_H=+1$.
	}
\label{Q_hel1_kvar}
\end{figure}

The effect of changing the magnetic field coherence scale is shown in Fig.~\ref{Q_hel1_kvar}
where the magnetic field strength and other parameters are fixed and only $k_{\rm mag}$
is varied. The magnetic field
with larger coherence length gives a larger signal, but there is a turning point as extremely large coherence scale fields will behave like uniform fields. The signal for smaller correlation length
is washed out but the suppression depends on the energy combination. This is to be expected
from the analysis of Ref.~\cite{Tashiro:2013bxa} since ${\overbar Q}$ with a certain energy 
combination is sensitive to the magnetic helicity power spectrum at a definite coherence scale 
that is determined by the combination of energies. To probe magnetic fields on small
length scales, it is necessary to consider gamma rays whose energies are close
together \cite{Tashiro:2013bxa}. Thus the energy bins also have to be smaller and this 
means that the statistics is poorer.

\section{The features at small $R$}
\label{sec:bump}

The Q-statistic is essentially a measure of the differential rotation found in the arrival direction of photons of 
different energies. $Q$ will be negative (positive) if the rotation is right (left)-handed as is depicted in Fig.~\ref{Qidea}.
The results of our Monte Carlo simulations, for example in Fig.~\ref{uniB_Q_100_hel1}, show that
${\overbar Q}$ has some oscillation at small $R$ which is made abundantly clear for the ${\overbar Q}$ using the bin with highest energies. 
Here we provide an explanation of this small $R$ feature.
What the Q-statistic allow us to probe is the shape of the PP surface. As already mentioned, it is clear from 
Fig.~\ref{nostoQexample} that high energy photons are found close to the LoS and the low energy ones 
are further out. However there is another important piece of information, namely
 the $z$ coordinate of their PP locations.

Remember that to give a contribution to $Q$, one requires a high energy photon ($\gamma_{HE}$) 
and a low energy photon ($\gamma_{LE}$) as is shown in Fig.~\ref{Qidea}. If most $\gamma_{LE}$ photons have PP locations
higher up on the PP surface than the $\gamma_{HE}$'s,
the Q-statistic measures the twist of the PP surface as one traverses it from
bottom to the top. This is in contrast to whenever the low energy 
photons originate from PP locations close to Earth when compared to those of $\gamma_{HE}$. As the twist is parity odd, these two cases contribute to 
${\overbar Q}$ with opposite signs and is the reason for these oscillations.

We can understand this effect explicitly with a little more thought. 
When $R$ is very small, we expect to see events with small bending angles. These mainly occur when the
lepton upscatters a photon towards Earth early after it was pair produced. The small $R$
observed $\gamma_{HE}$ ($\gamma_{LE}$) photons must therefore have originated from TeV leptons with high (low)
energies. 
Because TeV gamma ray have a MFP that decreases with energy, we then expect the $\gamma_{LE}$
to be produced at PP locations closer to Earth than the $\gamma_{HE}$ ones, 
therefore $Q$ initially measures the twist from top to bottom. Assuming that the helicity of the magnetic field is right-handed, the value of Q will become increasingly negative as $R$ departs from 0.

\begin{figure}
	\centering
	\includegraphics[width=0.75\textwidth]{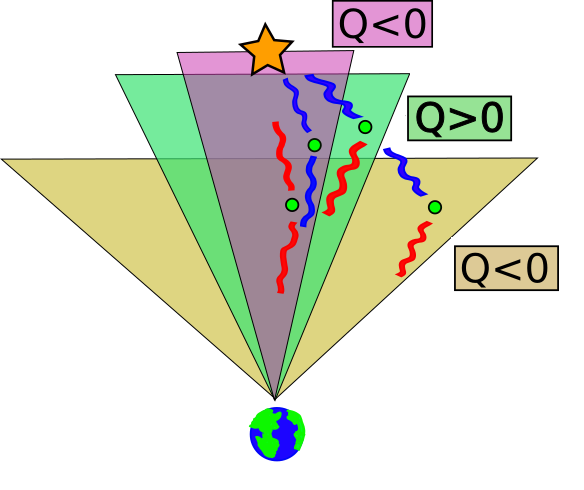} 
	\caption{The sign of the $Q$-statistic when applied to a halo produced by a magnetic field with $f_H=1$.
Here the squiggly lines represents photons of high (blue) and low (red) energies. The top photons are the initial
TeV gamma rays, the green points are the PP locations and the bottom photons are the upscattered GeV photons that 
are eventually observed. At small $R$, the events with small bending allows the $Q$-statistic to measure the twist of the PP surface from 
top to bottom (purple shaded region, $Q<0$). When $R$ gets larger
and PP surface region near the blazar enters the field of view, the new events entering the field of view contribute to 
$Q$ with the opposite sign as they are located further up the PP surface than the
high energy events which only occurs close to the LoS (green shaded region, $Q > 0$). Finally as $R$ gets large and the whole halo is exposed, 
the overwhelming low energy events at low $z$ and far from the LoS dominate the signal and drive $Q$ toward 
its asymptotic value (brown shaded region, $Q<0$).}
	\label{bumpskymap}
\end{figure}

However as $R$ increases further, two new things occur. First, the observed photons entering the field of view come
from events who experiences more bending. This means that the lepton had to travel further and in the process lost more energy, therefore the maximal energy of the new photons will be lower than those observed at small $R$ and won't contribute to the highest energy bin.
Hence the average $z$ of the PP location of the high energy photons is determined by the events at small $R$. 
Second, most of the volume entering the field of view will be located closer to the source due to 
projection effects; the observed volume is shown by the green cone of Fig.~\ref{bumpskymap}. We can then statistically 
expect the newly observed photons to have PP locations located near $z=0$. We are now in the reverse situation: a large majority of photons $\gamma_{LE}$ have PP locations that
are higher than the ones for the $\gamma_{HE}$. These give a contribution to $Q$ with the opposite sign, pulling the its value toward 0
 and sometime even changes its overall sign (compare the top and bottom panel of Fig.~\ref{uniB_Q_100_hel1}). 
 Finally when $R$ becomes large, 
the low energy events with PP locations at small $z$ and far from the LoS start contributing to $Q$ and 
dominate. This process, depicted in Fig.~\ref{bumpskymap}, is responsible for the features before the 
value of $Q$ flattens out to its asymptotic value.

\section{Conclusions}
\label{conclusions}

We have studied the effect of stochastic inter-galactic magnetic fields on the morphology 
of gamma ray halos. The dependence of the morphology on the magnetic field strength,
the coherence length, and the helicity were investigated. Most importantly, we have provided
an understanding of the structure of the halo in geometrical terms, as arising due to
the ``PP surface'' as determined by the magnetic field. In simple cases, the PP surface
can be found analytically (for example, Eq.~(\ref{analyticPP})).

To analyze the halo morphology, we have proposed a sharper version of the
Q-statistic in Eq.~(\ref{Qdefn}) and applied it to simulated halos. Our key finding is
that $Q$ is a powerful diagnostic of the magnetic helicity (Fig.~\ref{uniB_Q_100_hel1}), field 
strength (Fig.~\ref{Q_Bvar_k0d01_fh1_nolegend}) and 
coherence scale (Fig.~\ref{Q_hel1_kvar}). 
Based on the analytical work of Ref.~\cite{Tashiro:2013bxa}, we expect the sensitivity of $Q$ 
to the coherence
scale to depend crucially on the energies of the gamma rays that are used.
It would be interesting to quantitatively examine how the sensitivity of $Q$ to the coherence 
scale can be improved with a choice of energy bins.

In addition, our Monte Carlo simulations have revealed a bump in ${\overbar Q}(R)$
at small values of $R$ (see Fig.~\ref{uniB_Q_100_hel1}). We have understood and explained
this feature in terms of the PP surface in Sec.~\ref{sec:bump}. This new feature may
become useful in the analysis of real data in the future.

Our present study is limited in a few ways that we plan to overcome in future work. 
First, we have not included any background gamma rays. These will introduce noise in the
evaluation of ${\overbar Q}$ and the error bars will increase. We have also limited 
ourselves to stochastic isotropic magnetic fields but with only one $|{\bf k}|-$mode.
This is useful at this stage as it allows us to diagnose the effects of changing
the coherence scale. In future, we plan to include a spectrum of the magnetic
field as motivated by current observations \cite{Vachaspati:2016xji}.
In future we also plan to incorporate the full development of the electromagnetic cascade 
into our numerical code, perhaps along the lines of Ref.~\cite{AlvesBatista:2016urk} or \cite{Fitoussi:2017ble}. Once 
we have understood individual blazar halos, we will apply our techniques to the diffuse 
gamma ray background which is expected to contain halos due to unseen blazars as
well as those due to identified blazars.

\section*{Acknowledgement}
We thank Kohei Kamada and Andrew Long for comments.
TV is supported by the U.S. Department of Energy, Office of High Energy Physics, 
under Award No. DE-SC0013605 at ASUs.

\appendix
\section{Generation of Isotropic Random Magnetic Fields}
\label{Bgeneration}

To generate helical magnetic fields, we first decompose the magnetic field ${\bf B}({\bf x})$ in circularly polarized
modes with basis vectors ${\bf K}^{\pm}({\bf k})$ that are divergence-free eigenfunctions of the Laplace operator
\be
{\bf K}^{\pm}({\bf k})={\bf e}^{\pm}({\bf k})\text{e}^{i {\bf k} \cdot {\bf x}}
\equiv \frac{{\bf e_1}({\bf k})\pm i {\bf e_2}({\bf k})}{\sqrt{2}}\text{e}^{i {\bf k} \cdot {\bf x}}.
\ee
The triad of unit vectors, $\{ {\bf e_1},{\bf e_2},{\bf e_3} \}$, is constructed as
\be
{\bf e}_1\equiv\frac{{\bf n}_0\times \hat{\bf k}}{|{\bf n}_0\times \hat{\bf k}|},~
{\bf e}_2\equiv\frac{\hat{\bf k}\times {\bf e}_1}{|\hat{\bf k}\times {\bf e}_1|},~
{\bf e}_3=\frac{\bf k}{k}\equiv \hat{\bf k}
\ee
where ${\bf n}_0$ is any chosen unit vector such that ${\bf n}_0 \ne {{\bf {\hat k}}}$.

With these definitions, the ${\bf e}$'s form a right-handed orthonormal system and we have,
\be
\nabla\cdot {\bf K}^{\pm}=0,~\nabla\times {\bf K}^{\pm}=\pm k{\bf K}^{\pm},~
{\bf K}^{\pm *}({\bf k})= -{\bf K}^\pm(-{\bf k})
\ee

Hence any magnetic field can be decomposed as,
\begin{eqnarray}
{\bf B}({\bf x})&&=\int \frac{d^3 k}{(2\pi)^3} {\bf b}({\bf k}) \text{e}^{i {\bf k} \cdot {\bf x}}\nonumber\\
&&=\int \frac{d^3 k}{(2\pi)^3}\big[ b^+{\bf K}^{+}+b^- {\bf K}^{-}\big],
\end{eqnarray}
with the condition that 
\be
b^{\pm *}({\bf k})=-b^\pm(-{\bf k})
\label{realB}
\ee
to ensure that ${\bf B}({\bf x})$ is real. The divergence-free condition, $\nabla\cdot  {\bf B}=0$ , is 
automatically satisfied in this procedure.

We are interested in generating random magnetic fields with given energy ($E_B(k)$)
and helical ($H_B(k)$) power spectra. The relations between the modes $b^\pm ({\bf k})$ 
and the power spectra are given by
\be\label{averagebsquare}
\frac{1}{8\pi}\langle |{\bf B}({\bf x})|^2 \rangle =
\int \frac{k^2dk}{16\pi^3} \Big[ |b^+|^2+|b^-|^2 \Big ]
\equiv \int E_B(k) d \ln(k)
\ee
and 
\be\label{adotb}
\langle{\bf A}({\bf x})\cdot {\bf B}({\bf x}) \rangle=
\int \frac{kdk}{2\pi^2}\Big[ |b^+|^2-|b^-|^2 \Big ]
\equiv \int H_B(k) d \ln(k).
\ee

The ratio of $E_B$ and $H_B$ will be written in terms of a function $f_H(k)$ as \cite{Brandenburg:2004jv},
\be
H_B(k)=f_H(k)\frac{8\pi}{k}E_B(k),
\label{defnfH}
\ee
and the ``realizability condition'' leads to the restriction
\be
 -1\leq f_H(k) \leq 1
 \ee 
The field is non-helical if $f_H=0$, maximally right-handed if $f_H=+1$, and maximally left-handed
if $f_H=-1$. 

Eqs.~(\ref{averagebsquare}), (\ref{adotb}) and (\ref{defnfH}) allow us to write,
\be
|b^\pm|^2=\Big(\frac{2\pi}{k}\Big)^3[1\pm f_H(k)] E_B(k).
\ee
Hence the modes $|b^\pm ({\bf k})|$ are drawn from a normal distribution with mean 
$\mu^\pm=0$ and standard deviation $\sigma^\pm = (1\pm f_H)(2\pi/k)^3 E_B(k)$.
We then include a uniformly drawn phase angle $\theta^\pm({\bf k}) \in [0,2\pi )$ which yields, 
\be
b^\pm ({\bf k})=|b^\pm ({\bf k})| e^{i \theta^\pm({\bf k})}
\ee

In this paper we focus on stochastic magnetic fields that are {\it isotropic} but have power on a single
length scale $\lambda_c=2\pi/k_{\rm mag}$ and that have either $f_H(k)=0$ or $f_H(k)=\pm 1$. This 
corresponds to a delta function distribution for $E_B(k)$ and vanishing or maximal helicity of either
sign. To ensure that the magnetic fields are stochastically isotropic, we choose $N^2+1$ vectors 
${\bf k}_n$ ($n=1,...,N^2+1$) that discretize half of the two-sphere of directions in ${\bf k}-$space,
\begin{equation}
\label{kmode}
{\bf k}_n = k_{\rm mag}(\sin\theta_i \cos\phi_j,\sin\theta_i\sin\phi_j,\cos\theta_i),
\end{equation}
with
\begin{equation}
\theta_i=\cos^{-1}\Big( \frac{2i-1}{N}-1\Big), \ 
\phi_j=2\pi \frac{(j-1)}{N},
\end{equation}
for $i,j = 1,\ldots ,N$, and
\begin{equation}
k_{N^2+1}=k_{\rm mag}(0,0,1)
\end{equation}

Once we have ${\bf k}$ and $b^\pm ({\bf k})$ as described above, we compute
\be
{\bf b}({\bf k})= b^+{\bf K}^{+}+b^- {\bf K}^{-}
\ee
for every ${\bf k}={\bf k}_n$. We also find ${\bf b}(-{\bf k})$ using the reality condition
\be
{\bf b}(-{\bf k}) = {\bf b}^*({\bf k}).
\ee
Finally we obtain the random magnetic field,
\be\label{rmfonemode}
{\bf B}(\bf {x})=\frac{1}{(2N^2+2)}\sum_{{\bf k}\in K} {\bf b}({\bf k})e^{i {\bf k}\cdot {\bf x}}
\ee
where $K$ is the set of vectors $\{ {\bf k}_n, -{\bf k}_n \}$ for $n=1,\ldots ,N^2+1$.


\begin{thebibliography}{9}

\bibitem{Neronov:1900zz}
  A.~Neronov and I.~Vovk,
  Science {\bf 328}, 73 (2010)
  doi:10.1126/science.1184192
  [arXiv:1006.3504 [astro-ph.HE]].

\bibitem{Ando:2010rb}
  S.~Ando and A.~Kusenko,
  Astrophys.\ J.\  {\bf 722}, L39 (2010)
  doi:10.1088/2041-8205/722/1/L39
  [arXiv:1005.1924 [astro-ph.HE]].

\bibitem{Essey:2010nd}
  W.~Essey, S.~Ando and A.~Kusenko,
  Astropart.\ Phys.\  {\bf 35}, 135 (2011)
  doi:10.1016/j.astropartphys.2011.06.010
  [arXiv:1012.5313 [astro-ph.HE]].

\bibitem{Tashiro:2013ita}
  H.~Tashiro, W.~Chen, F.~Ferrer and T.~Vachaspati,
  Mon.\ Not.\ Roy.\ Astron.\ Soc.\  {\bf 445}, no. 1, L41 (2014)
  doi:10.1093/mnrasl/slu134
  [arXiv:1310.4826 [astro-ph.CO]].
  
\bibitem{Chen:2014qva}
  W.~Chen, B.~D.~Chowdhury, F.~Ferrer, H.~Tashiro and T.~Vachaspati,
  Mon.\ Not.\ Roy.\ Astron.\ Soc.\  {\bf 450}, no. 4, 3371 (2015)
  doi:10.1093/mnras/stv308
  [arXiv:1412.3171 [astro-ph.CO]].

\bibitem{Chen:2014rsa}
  W.~Chen, J.~H.~Buckley and F.~Ferrer,
  Phys.\ Rev.\ Lett.\  {\bf 115}, 211103 (2015)
  doi:10.1103/PhysRevLett.115.211103
  [arXiv:1410.7717 [astro-ph.HE]].

\bibitem{Finke:2015ona} 
  J.~D.~Finke, L.~C.~Reyes, M.~Georganopoulos, K.~Reynolds, M.~Ajello, S.~J.~Fegan and K.~McCann,
  Astrophys.\ J.\  {\bf 814}, no. 1, 20 (2015)
  doi:10.1088/0004-637X/814/1/20
  [arXiv:1510.02485 [astro-ph.HE]].

   \bibitem{Brandenburg:2004jv} 
  A.~Brandenburg and K.~Subramanian,
  Phys.\ Rept.\  {\bf 417}, 1 (2005)
  doi:10.1016/j.physrep.2005.06.005
  [astro-ph/0405052].
  
\bibitem{Durrer:2013pga} 
  R.~Durrer and A.~Neronov,
  Astron.\ Astrophys.\ Rev.\  {\bf 21}, 62 (2013)
  doi:10.1007/s00159-013-0062-7
  [arXiv:1303.7121 [astro-ph.CO]].


\bibitem{Wagstaff:2014fla} 
  J.~M.~Wagstaff and R.~Banerjee,
  JCAP {\bf 1601}, 002 (2016)
  doi:10.1088/1475-7516/2016/01/002
  [arXiv:1409.4223 [astro-ph.CO]].
  
\bibitem{Vachaspati:2016xji} 
  T.~Vachaspati,
  arXiv:1606.06186 [astro-ph.CO].


\bibitem{Elyiv:2009bx} 
  A.~Elyiv, A.~Neronov and D.~V.~Semikoz,
  Phys.\ Rev.\ D {\bf 80}, 023010 (2009)
  doi:10.1103/PhysRevD.80.023010
  [arXiv:0903.3649 [astro-ph.CO]].

\bibitem{Long:2015bda} 
  A.~J.~Long and T.~Vachaspati,
  JCAP {\bf 1509}, no. 09, 065 (2015)
  doi:10.1088/1475-7516/2015/09/065
  [arXiv:1505.07846 [astro-ph.CO]].
 
 \bibitem{AlvesBatista:2016urk} 
  R.~Alves Batista, A.~Saveliev, G.~Sigl and T.~Vachaspati,
  Phys.\ Rev.\ D {\bf 94}, no. 8, 083005 (2016)
  doi:10.1103/PhysRevD.94.083005
  [arXiv:1607.00320 [astro-ph.HE]].
  
  \bibitem{Broderick:2016akd} 
  A.~E.~Broderick, P.~Tiede, M.~Shalaby, C.~Pfrommer, E.~Puchwein, P.~Chang and A.~Lamberts,
  arXiv:1609.00387 [astro-ph.HE].
  
  \bibitem{Fitoussi:2017ble} 
T.~Fitoussi, R.~Belmont, J.~Malzac, A.~Marcowith, J.~Cohen-Tanugi and P.~Jean,
doi:10.1093/mnras/stw3365
arXiv:1701.00654 [astro-ph.HE].

 \bibitem{Tashiro:2013bxa} 
  H.~Tashiro and T.~Vachaspati,
  Phys.\ Rev.\ D {\bf 87}, no. 12, 123527 (2013)
  doi:10.1103/PhysRevD.87.123527
  [arXiv:1305.0181 [astro-ph.CO]].

  \bibitem{Ade:2015xua} 
  P.~A.~R.~Ade {\it et al.} [Planck Collaboration],
  Astron.\ Astrophys.\  {\bf 594}, A13 (2016)
  doi:10.1051/0004-6361/201525830
  [arXiv:1502.01589 [astro-ph.CO]].

\bibitem{Neronov:2009gh}
A. Neronov and D. Semikoz, 
Phys.Rev. D80 (2009) 123012, arXiv:0910.1920.

\bibitem{Ackermann:FS2}
M.~Ackermann {\it et al.} [Fermi-LAT Collaboration],
Astrophys.\ J.\ Suppl.\  {\bf 222}, no. 1, 5 (2016)
doi:10.3847/0067-0049/222/1/5
[arXiv:1508.04449 [astro-ph.HE]]. 
  
  \bibitem{TheFermi-LAT:2015ykq} 
  M.~Ackermann {\it et al.} [Fermi-LAT Collaboration],
  Phys.\ Rev.\ Lett.\  {\bf 116}, no. 15, 151105 (2016)
  doi:10.1103/PhysRevLett.116.151105
  [arXiv:1511.00693 [astro-ph.CO]].
  


  
\end{thebibliography}
\end{document}